\documentclass[12pt,a4paper,oneside]{article}
\usepackage{graphicx,amssymb,amsmath,color}
\graphicspath{{figures/}}
\usepackage{float}
\usepackage{bm}
\usepackage{empheq, cases}
\usepackage[linkcolor={black},citecolor={red},colorlinks=true,pagebackref=true,linktoc=all]{hyperref}
\usepackage{cite}
\usepackage{enumerate}
\usepackage[margin=2 cm]{geometry}
\usepackage{slashed}
\usepackage{multirow}
\usepackage[normalem]{ulem}
\usepackage[dvipsnames, x11names]{xcolor}
\usepackage{xparse}
\usepackage[most]{tcolorbox} 
\usepackage[compat=1.0.0]{tikz-feynman}
\usepackage[labelfont=bf]{caption}
\usepackage{subcaption}

\usepackage{cleveref}
\usepackage{cancel}
\usepackage{dsfont}

\definecolor{flag}{rgb}{1.0, 0.03, 0.0}

\def\beq{\begin{equation}}
\def\eeq{\end{equation}}
\def\bea{\begin{eqnarray}}
\def\eea{\end{eqnarray}}

\def\bit{\begin{itemize}}
\def\eit{\end{itemize}}

\def\l{\left}
\def\r{\right}

\def\baa{\begin{array}}
\def\eaa{\end{array}}

\def\d{\partial}

\def\simgt{\mathrel{\lower2.5pt\vbox{\lineskip=0pt\baselineskip=0pt
           \hbox{$>$}\hbox{$\sim$}}}}
\def\simlt{\mathrel{\lower2.5pt\vbox{\lineskip=0pt\baselineskip=0pt
           \hbox{$<$}\hbox{$\sim$}}}}

\def\bfc{\begin{figure}\begin{center}}
\def\efc{\end{center}\end{figure}}
\def\nn{\nonumber\\}

\definecolor{chromeyellow}{rgb}{1.0, 0.65, 0.0}
\definecolor{darkcoral}{rgb}{0.8, 0.36, 0.27}
\definecolor{cadmiumgreen}{rgb}{0.0, 0.42, 0.24}

\renewcommand{\Im}{\operatorname{Im}}
\newcommand{\abs}[1]{\left\lvert#1\right\rvert}
\newcommand{\quotes}[1]{``#1''}

\NewDocumentEnvironment{trungenv}{+b}{%
  \noindent\textcolor{blue}{\bfseries\# [Trung] #1}%
}{}

\let\oldbm\bm
\renewcommand{\bm}[1]{\oldbm{\mathrm{#1}}}

\hypersetup{
bookmarksnumbered=true,
bookmarksopen=true,
bookmarksopenlevel=1,
linktoc=all,
hypertexnames=true,
colorlinks=true,
pdftoolbar=true, 
pdfmenubar=true, 
pdfstartview={FitV}, 
pdfpagemode=UseOutlines,
linkbordercolor=red,
citecolor = Red, 
filecolor = magenta, 
linkcolor = blue,
urlcolor = green 
}

\begin{document}

\begin{flushright}
\hspace{3cm} 
SISSA  21/2024/FISI
\end{flushright}

\vspace{.6cm}

\begin{center}
{\Large \bf Q-ball perturbations with more details: linear analysis vs lattice}
\vspace{0.5cm}

\vspace{1cm}{Aleksandr Azatov$^{a,b,c,1}$, Quoc Trung Ho$^{a,b,2}$, Mohamed Mahdi Khalil$^{a,b,3}$ }
\\[7mm]
 {\it \small
$^a$ SISSA International School for Advanced Studies, Via Bonomea 265, 34136, Trieste, Italy\\
$^b$ INFN - Sezione di Trieste, Via Bonomea 265, 34136, Trieste, Italy\\[0.1cm]
$^c$ IFPU, Institute for Fundamental Physics of the Universe, Via Beirut 2, 34014 Trieste, Italy\\[0.1cm]
}
\end{center}

\bigskip \bigskip \bigskip

\centerline{\bf Abstract} 
  \begin{quote}
We analyze in detail the interactions between non-topological soliton (Q-ball) and its perturbations. We extend the previous literature by carefully identifying the domain of applicability of linear analysis as well discussion of the FLS Q-balls. Applications to the early universe physics are briefly commented.
\end{quote}

\vfill
\noindent\line(1,0){188}
{\scriptsize{ \\ E-mail:
\texttt{$^1$\href{mailto:ZZZ@NOSPAMsissa.it}{aleksandr.azatov@sissa.it}}, 
\texttt{$^2$\href{mailto:quoctrung.ho@NOSPAMsissa.it}{quoctrung.ho@sissa.it}},
\texttt{$^3$\href{mailto:XXX@NOSPAMsissa.it}{mkhalil@sissa.it}}
}
}

\newpage
\tableofcontents

\section{Introduction}
Non-topological solitons \cite{Coleman:1977py,Friedberg:1976ay,
Friedberg:1976az,
Friedberg:1976me} (for reviews, see \cite{Lee:1991ax,Nugaev:2019vru}) are very interesting objects which can emerge in various theories beyond the Standard Model. 
A special class of these objects, commonly referred to as Q-balls, can form if there is a complex field charged under conserved a $U(1)$ symmetry and there exists a field configuration with charge $Q$, which has less energy than $Q$ quanta of the free particles: 
\bea
E^Q_{\text{Q-ball}} < Q m_\Phi.
\eea
Given the charge conservation and the energy considerations, such configurations are  stable.
These objects have garnered significant interest within particle physics  phenomenology community due to their potential applications in dark matter model building and their possible connection to the baryon asymmetry of the universe
\cite{Kusenko:1997si,Kusenko:1997ad,Enqvist:1998xd,Kusenko:2001vu,
Roszkowski:2006kw,
Shoemaker:2009kg,
Kasuya:2011ix,
Kasuya:2012mh,
Kawasaki:2019ywz,Ponton:2019hux,Krylov:2013qe,Bishara:2017otb}. 
Unlike topological solitons, the production of Q-balls does not occur via the Kibble mechanism. 
Instead, it can take place during first- or second-order phase transitions, where regions with net non-zero charge are compressed by bubble walls (or domain walls) \cite{Frieman:1988ut, Griest:1989cb}, and this process is referred to as solitogenesis.
Later, the Q-ball distribution can evolve further by accreting free particles from surrounding plasma \cite{Griest:1990kh,Postma:2001ea,Frieman:1988ut} in a  process dubbed solitosynthesis. 
A precise calculation of this process requires detailed knowledge of Q-ball interaction with the surrounding plasma particles, particularly with the quanta of the $\phi$-field that form the Q-balls. 
The analysis of these interactions will be the primary focus of this paper.

Recently the Ref.\cite{Saffin:2022tub,Cardoso:2023dtm} 
have analyzed the interactions of the Q-ball with its perturbations, focusing on the processes when the Q-ball energy can be extracted.
In our paper we extend these results in the following directions: 
\bit
	\item We study in detail the applicability of the linear treatment of the perturbations by making comparison with lattice simulation, and highlight the possible applications for the process of solitosynthesis.
	\item We provide an intuitive understanding of the energy extraction process by matching energy of the perturbations to the change of the Q-ball self-energy.
    
	\item We provide the analysis of the energy extraction process for the Q-ball model with two fields (the simplest UV complete lagrangian) both in linear and non-linear regimes.
\eit

The paper is organized as follows: in the Section.~\ref{sec:1field} we start by discussing the perturbations of the Q-ball for the model with one 
complex field.
In the Section.~\ref{sec:2field} we discuss the perturbations for the model with one complex and one real field. 
Both sections are divided in subsections discussing peculiarities of the linear and non-linear analysis. 
We conclude by summarizing the main results in  Section.~\ref{sec:summary}.

\section{One-field Q-balls}
\label{sec:1field}
\begin{figure}[H]
	\centering
\includegraphics[width={0.7\textwidth}]{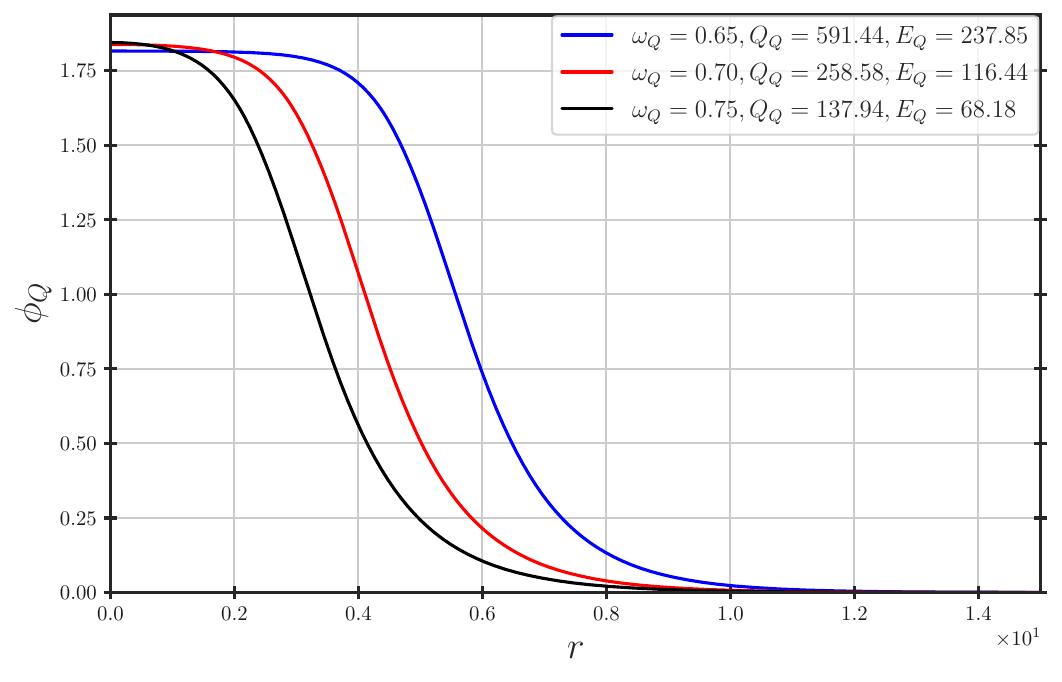}
	\caption{The Q-ball profiles with varying internal frequency in $D=3$ spatial dimension, with self-coupling $g = 1/3$. The total charge $Q_Q$ and energy $E_Q$ are reported to verify the stability of the Q-ball against decay into free particles, $E_Q < Q_Q m_\Phi$.}
	\label{fig: qball onefield}
\end{figure}

Let us start by reviewing the Q-ball solutions and their perturbations following the recent literature \cite{Saffin:2022tub, Cardoso:2023dtm}. 
Both of the mentioned references have considered  the model with one complex scalar field with the action given by:
\bea
\label{eq:1-field-action}
&&S[\Phi] = \int\mathrm{d}^4x\;\l[ |\partial_\mu \Phi|^2 - V \r],\qquad V = \mu^2\,|\Phi|^2 - \lambda\,|\Phi|^4+g\,|\Phi|^6 ,
 \eea
where we have used mostly negative metric $\eta_{\mu\nu}=\rm{diag}(1,-1,-1,-1)$.
This action can be written in terms of dimensionless quantities if we perform the transformations $ x \to x/\mu, \,\Phi \to \mu\,\Phi / \sqrt{\lambda},\, g \to \lambda^2\,g/\mu^2 $, 
\bea
\label{eq:ac-dimless}
&&S[\Phi] = \frac{1}{\lambda}\int\mathrm{d}^4x\;\l[ |\partial_\mu \Phi|^2 - V\r],\quad\qquad V = |\Phi|^2 - |\Phi|^4+g\,|\Phi|^6.   
\eea
We will use the action in this form for the rest of this section. 
The Q-ball is a time-dependent classical solution of the equations of motion of the following form: 
\bea
\Phi_Q(\mathbf{x}, t)=\frac{\phi_Q(r)}{\sqrt{2}} e^{-i \omega_Q t},
\eea
where the function $\phi_Q$ satisfies the equation
\bea
\phi_Q''(r) + \frac{2}{r} \phi_Q'(r) + \dfrac{1}{2} \,\omega_Q^2\phi_Q(r) - \frac{\d V}{\d \Phi^\ast}\Big|_{\Phi =\phi_Q/\sqrt{2}} = 0,
\eea
and $~'~$ stands for derivative in radial direction.
We have assumed for simplicity that the solution has zero angular momentum (non-rotating Q-ball).
The function $\phi_Q$ satisfies  the boundary conditions $ \phi_Q'(r=0) = 0$ (requirement of regularity at the origin), and $\phi_Q(r\to\infty) = 0 $ (requirement of finite energy). 
For the quantum stability, the Q-ball must satisfy:
\bea
\int_0^{\infty} dr\, r^2 \l( \frac{1}{2}\big(\phi_Q'^{\,2}+\omega_Q^2 \phi_Q^2\big) +V(\phi_Q)\r) < \int_0^\infty dr\, r^2\, \omega_Q\phi_Q^2 .
\eea
We show the solutions for vaious Q-ball charges on Fig.~\ref{fig: qball onefield}.

Let us proceed  to the perturbations  of the Q-ball $\Phi = \Phi_Q + \Phi_1$, then it is easy to show that the perturbation  $\Phi_1$\cite{Smolyakov:2017axd} satisfies the equations 
\begin{align}
	&\square \Phi_1 + U(r) \Phi_1 +  W(r) \Phi_1^\ast\,e^{-2i \omega_Q t} = 0, \label{eq: EoM perturbation Q-ball}
\end{align}
where $U$ and $W$ are defined as
\bea	
U= 1-2 \phi_Q^2+\frac{9 g}{4}\phi_Q^4, \qquad
W= -\phi_Q^2+\frac{3 g}{2}\phi_Q^4.
\eea
Note that the last term of Eq.~\eqref{eq: EoM perturbation Q-ball} introduces mode-mixing and consequently no monochromatic solutions exist.
In general solutions to these equations can be parameterized in the following way  \cite{Ciurla:2024ksm}:
\bea
\Phi_1 = \eta_+(\mathbf{x})\,e^{-i\omega_+t} + \eta_-(\mathbf{x})\,e^{-i\omega_-t}, \qquad
\omega_\pm = \omega_Q\pm \omega.
\eea
Depending on the value of the $\omega$ parameter we can classify these perturbations as follows \cite{Ciurla:2024ksm}\footnote{The only exceptions to this parametrization are the modes corresponding to the Lorentz boost and the Q-ball charge change \cite{Smolyakov:2017axd,Smolyakov:2019cld,Ciurla:2024ksm}.}
\bit
\item zero modes $\omega=0$
\item bound modes $|\omega_\pm| < 1$ 
\item half propagating modes $|\omega_+|>1,|\omega_-|<1$ or vice versa
\item propagating modes $|\omega_\pm|>1$
\eit
In this paper we are interested in understanding the Q-ball interaction with the plasma so we will focus only on the propagating and half propagating modes. 
The equation of motion for the $\eta_\pm(r)$ functions then becomes:
\bea
\label{eq:1fieldpert}
{\bf \nabla^2}\eta_{\pm} + \l[\omega_\pm^2-U\r]\eta_\pm-W \eta^\ast_\mp=0.
\eea
Far from the Q-ball, where $\phi_Q\to 0$, the term that mixes $\eta_+,\eta_-$ perturbations vanishes, and the equation of motion becomes the usual Klein-Gordon for the free scalar fields.
Interestingly as was shown in the original references \cite{Saffin:2022tub,Cardoso:2023dtm}, at the linear order in perturbations the following current is conserved:
\bea
\label{eq:jeta-cons}
\bm{J}_\eta \equiv 2 \Im \left( \eta_+^\ast\bm{\nabla}\eta_+ -\eta_-^\ast\bm{\nabla}\eta_- \right),
\eea
where in our notations the  bold letters, ${\bm \nabla \cdot \bm J}$,  denote the spatial vectors. Then obviously the divergence $\bm \nabla \cdot \bm J_\eta=0$, and the  corresponding flux through any closed surface must be equal to  zero.
We proceed by analysis of the S-wave scattering, which is sufficient to grasp the most important qualitative features of the Q-ball interaction with its perturbations and the discussion for non-spherical waves is presented in the Appendix.~\ref{app:non-sphere}.
For the S-waves the equation of motion becomes  
\bea
\label{eq:spherical-sym}
\eta''_{\pm}+\frac{2\eta'_{\pm}}{r}+\l[\omega_\pm^2-U\r]\eta_\pm-W \eta^\ast_\mp=0.
\eea
Far from the Q-ball the potential becomes constant, and consequently we obtain the solutions of free spherical waves
\begin{align}
	&\eta_{\pm}(r\to \infty)= \frac{1}{\sqrt { \abs{k_{\pm}}}\,r}\l[A_\pm^{\rm in} e^{-i k_\pm r} + A_{\pm}^{\rm out}e^{i k_\pm r}\r], \label{eq:asympt-S} \\
	&k_{\pm}=\pm \mathrm{sign}(\omega)\sqrt{(\omega_\pm^2-1)}= \mathrm{sign}(\omega_\pm)\sqrt{(\omega_\pm^2-1)}. \nonumber
\end{align}
The $\bm J_\eta$ current conservation imposes the following constraints on the amplitudes:
\bea
\label{eq:number-cons}
\boxed{
\abs{A^{\rm out}_+}^2+\abs{A^{\rm out}_-}^2=|A^{\rm in}_+|^2+|A^{\rm in}_-|^2.
}
\eea
Let us comment on the various symmetry properties of these perturbations. 
The system in Eq.~\eqref{eq:1fieldpert} is linear, which leads to the following relation between the expansion coefficients:
\bea
\vec{A}^{\rm out} = S \vec A^{\, \rm in}, \qquad
\vec{A}^{\, \rm out, in} = 
\l[\baa{c} 
A_+^{\, \rm out, in}\\
\left( A_-^{\, \rm out, in} \right)^\ast
\eaa\r]
\label{eq:S2}
\eea
The matrix $S$ must be unitary due to the Eq.~\eqref{eq:number-cons} and symmetric, since the Eq.~\eqref{eq:1fieldpert} is real (see for details and proof Appendix \ref{app:symmetry}).
Additionally, we note that exchanging $\omega \to -\omega$ simply flips the modes $\eta_+$ and $\eta_-$. 
This together with the symmetries of the matrix $S$ leads to the following relation
\bea
\label{eq:symm-amp}
\l|\frac{A_\pm^{\rm out}(\omega)}{A_-^{\rm in}(\omega)}\r|_{A_+^{\rm in}=0}=\l|\frac{A_\mp^{\rm out}(\omega)}{A_+^{\rm in}(\omega)}\r|_{A_-^{\rm in}=0},
\eea
and we refer for the derivation and more detailed discussion to the Appendix.~\ref{app:symmetry}.
Similarly to the current $\bm{J}_\eta$, we can also build the $U(1)$ charge and energy currents 
\begin{align}
\bm{J}_Q = (-i) \big[ \Phi_1^\ast\,\bm{\nabla}\,\Phi_1 - \Phi_1\,\bm{\nabla}\,\Phi^\ast_1 \big] \\
\bm{J}_E = (-1) \big[ \dot{\Phi}_1^\ast\,\bm{\nabla}\,\Phi_1 - \dot{\Phi}_1\,\bm{\nabla}\,\Phi_1^\ast \big].
\end{align}
In the asymptotic region, the average fluxes of charge and energy are
\begin{align}
\mathcal{F}_Q &= (4\pi)^{-1}\lim_{r\to\infty}\oint\mathrm{d}\bm{\Sigma}\cdot\langle\bm{J}_Q\rangle_{T \Omega} = 2 s_\omega\sum_{\pm} s \, \left( \abs{A_s^{\rm out}}^2 - \abs{A_s^{\rm in}}^2 \right), \label{eq: Q flux} \\
\mathcal{F}_E &= (4\pi)^{-1}\lim_{r\to\infty}\oint\mathrm{d}\bm{\Sigma}\cdot\langle\bm{J}_E\rangle_{T \Omega} = 2 s_\omega\sum_{\pm} s \, \omega_s \, \left( \abs{A_s^{\rm out}}^2 - \abs{A_s^{\rm in}}^2 \right), \label{eq: E flux}
\end{align}
where $\langle \dots \rangle_{T \Omega}$ denotes the average over sufficient long time and over all directions in space.

After this preliminary discussion we can proceed to the discussion of the charge and energy exchange between Q-ball and its perturbations.
Similarly to the Ref\cite{Cardoso:2023dtm}, we look at the relative difference between incoming and outgoing energy and charge fluxes by defining the quantities 
\begin{align}
	1 + Z_Q \equiv \frac{{\cal F}_Q^{\rm out}}{{\cal F}_Q^{\rm in}} = \abs{\frac{\abs{A_+^{\rm out}}^2-\abs{A_-^{\rm out}}^2}{\abs{A_+^{\rm in}}^2-\abs{A_-^{\rm in}}^2}}, \qquad
	1 + Z_E \equiv \frac{{\cal F}_E^{\rm out}}{{\cal F}_E^{\rm in}} = \abs{\frac{\omega_+ \abs{A_+^{\rm out}}^2-\omega_-\abs{A_-^{\rm out}}^2}{\omega_+ \abs{A_+^{\rm in}}^2-\omega_-\abs{A_-^{\rm in}}^2}}.
\label{eq:1fieldZEZQ}
\end{align}
Note that \cite{Saffin:2022tub} used another notion of amplification factors based on the charge/energy contained within a sphere far away from the source, instead of the fluxes through the surface of such a sphere.
In our paper, we only use the definition in Eq.~\eqref{eq:1fieldZEZQ}, hence for convenience the superscript $\mathcal{F}$ will be dropped from now on without causing confusion.
We can then consider the initial state where only one frequency mode is present (either $A_{\rm in}^{+}=0$ or $A_{\rm in}^{-}=0$). 
Then using Eq.~\eqref{eq:number-cons} we know that in the final states we will generically have both $\pm$ modes $A_\pm^{\rm out} \neq 0$, implying in principle there is energy and charge exchange between the Q-ball and the perturbations.

\begin{figure}[H]
	\centering
	\includegraphics[width={0.9\textwidth}]{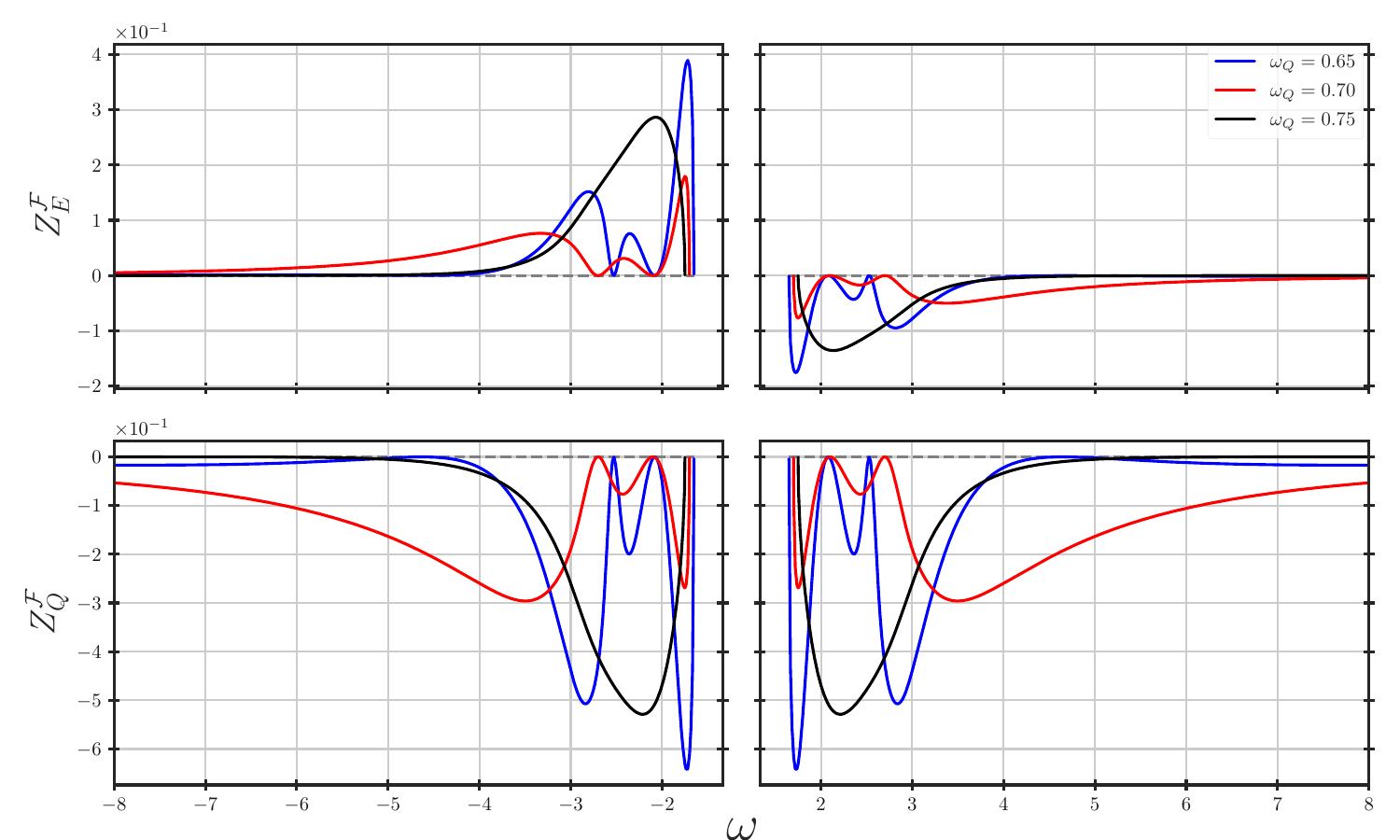}
	\caption{The relative amplification factors of incoming mode $\eta_+$ in terms of energy (top figure) and charge (bottom figure) of various Q-ball profiles shown at Fig.~\ref{fig: qball onefield}.
	We see that both $Z_{E, Q}$ tends to have more peaks for scattering off a thinner Q-ball (i.e smaller $\omega_Q$). 
	The results for incoming mode $\eta_-$ is obtained by exploiting the symmetry at Eq.~\eqref{eq:symm-amp}.
	}
	\label{fig: one field Z}
\end{figure}

The results for this quantities are reported in the Fig.~\ref{fig: one field Z}. 
Similarly to what was reported in the 2D case, there could be energy and charge exchange between the Q-ball and its perturbations.

\begin{figure}[H]
	\centering
	\includegraphics[width=\linewidth]{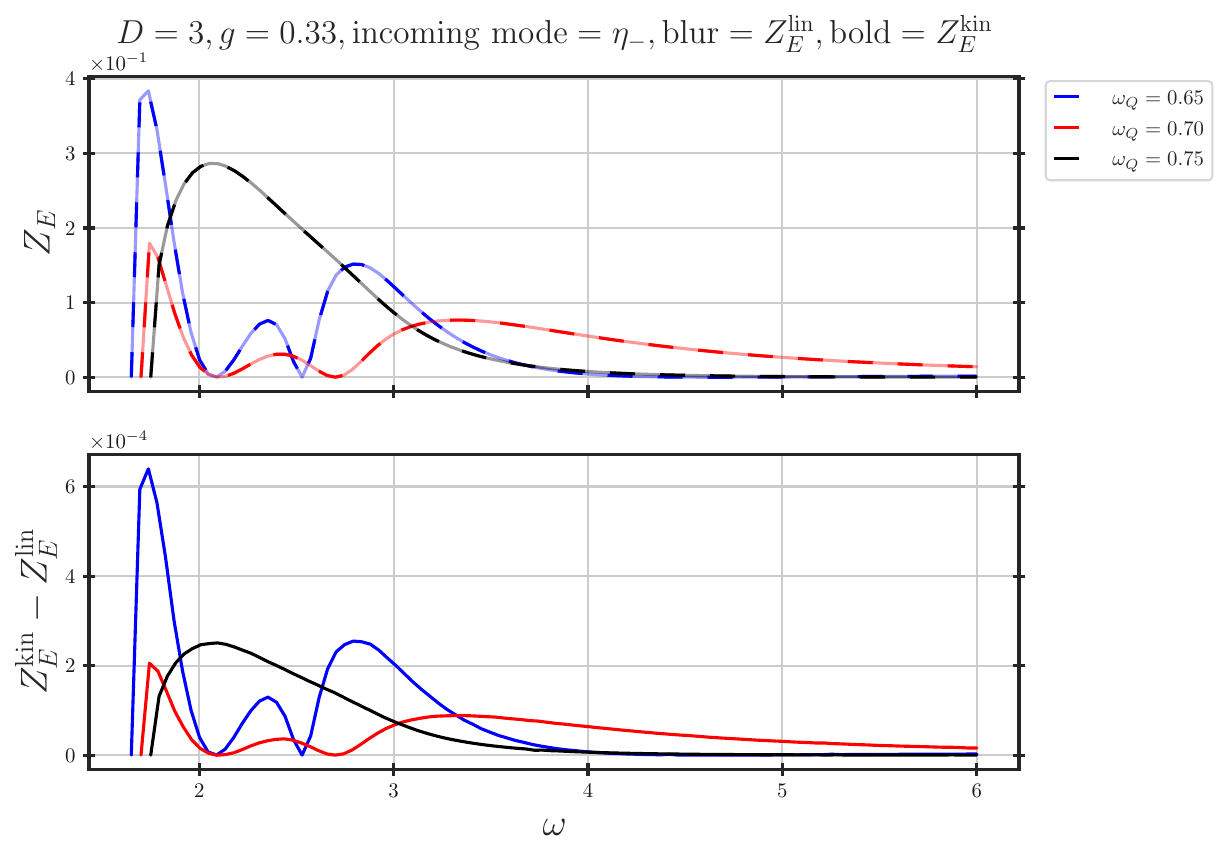}
	\caption{The kinematic and linear relative energy amplifications for an incoming $-$ mode (top figure) and their difference (bottom figure), here $Z_E^{\rm kin} \equiv \tilde{Z}_E$ and $Z_E^{\rm lin}$ is obtained using linear perturbation regime.
	It is obvious that the shape of $Z_E^{\rm kin} - Z_E^{\rm lin}$ matches the shape of $Z_E^{\rm kin}$ for each value of $\omega_Q$, as predicted by Eq.~\eqref{eq: ZE expansion}.
}
	\label{fig:ZE_check_2}
\end{figure}

This process becomes intuitively very clear once we quantize the perturbations and the quantities $|A_\pm^{\rm in,out}|^2$ are becoming the probabilities to find the corresponding particles and antiparticles in the initial and final states. Then the current conservation in Eq.\ref{eq:jeta-cons}-\ref{eq:number-cons} means that the total number of particles and antiparticles must be conserved, and we will call $\bf J_\eta$ particle number current.
The energy exchange with the Q ball at the linear level can only occur due to the following processes  (we are using the total $U(1)$ charge conservation )
\bea
&&Q+ \phi \to Q+\phi\nn
&&Q+ \phi^\dagger \to Q+\phi^\dagger\nn
&&Q+\phi \to (Q+2)+\phi^\dagger\nn
&&Q+\phi^\dagger \to (Q-2) +\phi,
\eea
where $(Q\pm2)$ stand for the Q-balls with the corresponding charges and $\phi,\phi^\dagger$ are the quanta of the field $\Phi$.
The first two processes are just elastic scatterings, so they cannot extract or absorb neither charge nor energy. 
To analyze the other two let us assume that Q-ball has positive charge $Q>0$, then in the process $Q+\phi \to (Q+2)+\phi^\dagger$ its charge is increased, so the energy will be increased as well, thus the energy is absorbed from the perturbation. 
Similarly the process $Q+\phi^\dagger \to (Q-2) + \phi$ leads to the decrease of the charge of the Q-ball and extraction of energy. 
From this we can immediately see that energy extraction happens only if the incoming particle has the charge opposite to the Q-ball charge. 
As a sanity check we can  calculate the energy amplification  from this considerations and compare it with Eq.~\eqref{eq:1fieldZEZQ}
\bea
\label{eq:ZEfromBE}
\tilde Z_E(A_+^{\rm in} = 0, A_-^{\rm in} = 1) = \frac{|A_+^{\rm out}|^2}{|\omega_-|} \big[ E(Q) - E(Q-2) \big],  \nn
\tilde  Z_E(A_+^{\rm in} = 1, A_-^{\rm in} = 0) = \frac{|A_-^{\rm out}|^2}{\omega_+} \big[ E(Q) - E(Q+2) \big],
\eea
where the factor $\abs{A_+^{\rm out}}^2$ takes into account the probability of the transition $Q+\phi^\dagger \to (Q-2)+\phi$. 
Comparing with the expression in Eq.~\eqref{eq:1fieldZEZQ} we can see that quantities indeed match approximately, 
\begin{align}
Z_E(A_+^{\rm in} = 0, A_-^{\rm in} = 1)
&= \dfrac{\omega_+ \abs{A_+^{\rm out}}^2 - \omega_- \abs{A_-^{\rm out}}^2}{\abs{\omega_-} } - 1
= \dfrac{2 \omega_Q}{\abs{\omega_-}} \abs{A_+^{\rm out}}^2 \nonumber \\
&=  \dfrac{2 (dE_Q/dQ_Q) \abs{A_+^{\rm out}}^2}{\abs{\omega_-}} 
\approx \dfrac{\abs{A_+^{\rm out}}^2}{\abs{\omega_-}} \l[E(Q) - E(Q-2)\r], 
\end{align}
where we have used the relations (see for example \cite{Lee:1991ax},\cite{Heeck:2020bau}):
\bea
\label{eq:EQder}
\frac{d E(Q)}{d Q}=\omega_Q,~~E(Q+2)-E(Q)=2 \omega_Q+ 2\frac{d^2 E_Q}{dQ^2}+...
\eea
It is clear that two expressions for the energy extraction are equal up to the second and higher derivative terms
\begin{align}
& Z_{E}(A_+^{\rm in} = 0, A_-^{\rm in} = 1)
	\simeq \tilde Z_{E}(A_+^{\rm in} = 0, A_-^{\rm in} = 1)\l(1-\frac{2\frac{d^2 E}{d Q^2}}{E(Q)-E(Q-2)}+ \dots \r).
	\label{eq: ZE expansion}
 \end{align}
For large values of the Q-ball charge the second term scales as $Q^{-1}$, and is strongly suppressed.
Interestingly the sign of it is always positive, since the Q-ball classical stability dictates \cite{Friedberg:1976me} that
\bea
\frac{d^2E_Q}{dQ^2}=\frac{d \omega_Q}{dQ} <0 .
\eea
We illustrate these relations on the Fig.~\ref{fig:ZE_check_2}, where we clearly see that the two results match and the difference is proportional to $Z_E$ as expected from Eq.~\eqref{eq:EQder}.

On the wave-equation side, the energy extraction can be seen from the following. 
Let us assume that $\omega_Q>0$ and $\omega>0$ (taking $\omega<0$ will just lead to interchange of $\eta_-\leftrightarrow \eta_+$), then we always have
\bea
|\omega_+|> |\omega_-|,
\eea
thus the energy of the $\eta_+$ mode is always larger than the energy of $\eta_-$ mode. Then in agreement with Eq.~\eqref{eq:ZEfromBE} we will have the energy extraction only when the $\eta_-$ (antiparticle) is in the initial state.

Similarly we can consider the scattering of the half-bound modes.
In this case only $\eta_+$ can propagate and from the current $\bm J_\eta$ conservation we immediately obtain
\bea
\abs{A_+^{\rm in}}^2=\abs{A_+^{\rm out}}^2,
\eea
so that the collision is always elastic and no energy and charge exchange with the Q-ball is possible.

\subsection{Q-ball evolution in the early universe}
\begin{figure}
	\begin{center}
		\includegraphics[scale=0.5]{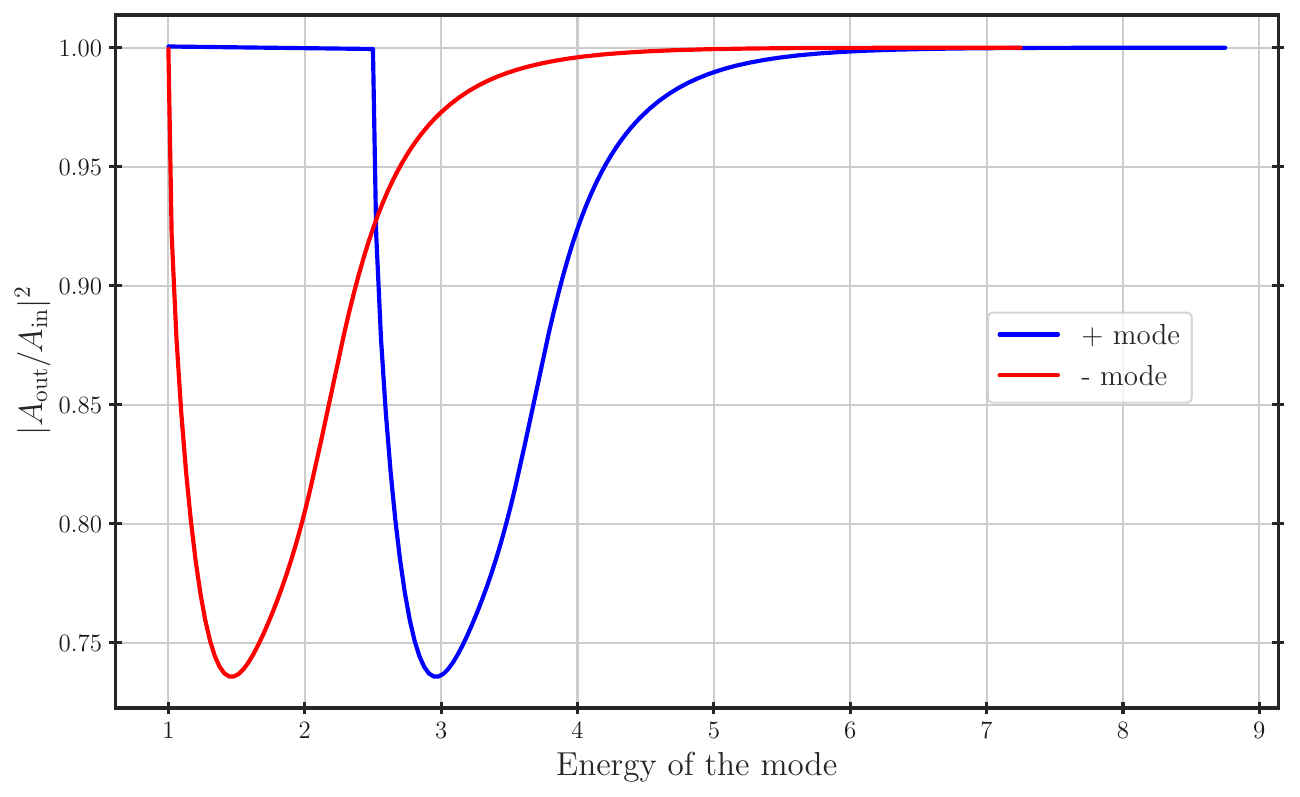}
    \end{center}
    \caption{Reflection probabilities for the $\pm$ modes $|A_\pm^{\rm out}/A_\pm^{\rm in}|^2$, with $g=1/3$ and $\omega_Q = 0.75$. There is always only one mode (either $+$ or $-$ in incoming state).
	}
	\label{Fig:reflection}
\end{figure}

The discussion of the linear perturbations and the selection rules for the  Q-ball and its perturbation interactions lead to a very interesting conclusion. 
Suppose Q-ball is formed and in the early universe and we are looking at its evolution due to the interactions with the surrounding plasma. 
If the plasma particles are energetic enough, both of the reactions
\bea
&&Q+\phi \to (Q+2)+\phi^\dagger\nn
&&Q+\phi^\dagger \to (Q-2) +\phi
\eea
are open, even though the rates are different (see Fig.~\ref{Fig:reflection}). 
Note that the blue curve is obtained by shifting the red one by the factor $2 \omega_Q$ which 
 follows from the symmetry in Eq.~\eqref{eq:symm-amp}. 
 We can see that at classical level once the energy of $\eta_+$ mode 
 becomes less than $\omega_+ < 1+2 \omega_Q$ corresponding $\eta_-$ 
becomes bounded, and classically we can have  only the following processes:
\bea
\label{eq:min-energy}
&&E < 1+2 \omega_Q\nn
&&Q+\phi^\dagger \to (Q-2) +\phi \hbox{~~open}\nn
&&Q+\phi  \to (Q+2) +\phi^\dagger\hbox {  ~~forbidden}.
\eea
Therefore, in this range of energies Q-balls can only lose their charge due to the interactions with plasma. 
Note that a very similar conclusion can be obtained just by considering the energy conservation conditions. Let us suppose that the $\phi$ particle has energy $E_*$ then
\bea
&&Q+\phi \to (Q+2)+\phi^\dagger \hbox{ is forbidden if} \nn
&&m_\Phi > E(Q)+E_*-E(Q+2).
\eea
Then the minimal energy when the inelastic reaction can occur is equal to:
\bea
E_*=E(Q+2)-E(Q)+m_\Phi=1+2 \omega_Q+2\frac{d^2 E}{dQ^2}+...
\eea
So we can see that up to higher derivative terms this expression matches with the expression in Eq.~\eqref{eq:min-energy}.

We remind the reader that these results have been obtained in the linear approximation and ignoring the quantum effects. 
The validity of the linear approximation will be discussed in the next session, however for sufficiently small perturbations we expect these results and in particular the conservation of the current Eq.~\eqref{eq:jeta-cons} to be exact. 
On the other hand, the quantum effects are obviously absent in this treatment of perturbations.
For example, if we add an interaction
\bea
\delta {\cal L}= \frac{1}{\Lambda_*^2}F_{\mu\nu}^2 |\phi|^2
\eea
to our lagrangian, the reaction
\bea
Q+\phi \to (Q+1)+\gamma\gamma
\eea
 will be open, however it will be suppressed compared to the \quotes{classical transition} by the coupling size and phase space factors.

\subsection{Non-linear regime}
So far our analysis was focused solely on the linear treatment of the perturbations. 
We proceed here with the analysis of the non-linearities by solving the system of differential equations on the lattice.

We do it by discretizing the spatial coordinate using 4th order finite difference, and evolve over time with Runge-Kutta 4th order method. 
We have used the absorbing boundary condition at the boundary corresponding to the spatial infinity. 
Implementing the  other boundary conditions (e.g Dirichlet) is also possible, as long the boundary is significantly far away so that the reflected waves do not spoil the measurements of fluxes.
The initial conditions of the system including the Q-ball background and a monochromatic Gaussian wavepacket that scattering off the Q-ball read
\begin{align}
	&\Phi(0, r) = \Phi_Q(0, r) + \delta \left( \dfrac{r_0}{r} \right) e^{- \frac{(r-r_0)^2}{2 \sigma_r^2}} e^{-i s_{\omega_0} \sqrt{\omega_0^2 - 1} r}, \\
	&\partial_t \Phi(0, r) = -i \bigl[ \omega_Q \Phi_Q(0, r) + \omega_0 (\Phi(0, r) - \Phi_Q(0, r)) \bigr].
\end{align}
The factor $\left( \dfrac{r_0}{r} \right)$ is negligible when the packet width $\sigma_r$ is relatively smaller than the distance from the wavepacket to the Q-ball $r_0$. 
This is not the case when $\omega_0$ is close to the threshold of having two propagating modes, i.e when one of the wavenumber $k_+$ or $k_-$ approaches $0$, and consequently the wavelength becomes significantly large.
In such a case, in order to capture enough oscillations to have good resolution to the mode frequencies, we need to have sufficiently large $\sigma_r$ and hence the asymptotic damping behavior becomes relevant.
\begin{figure}[H]
	\centering
	\includegraphics[width={\textwidth}]{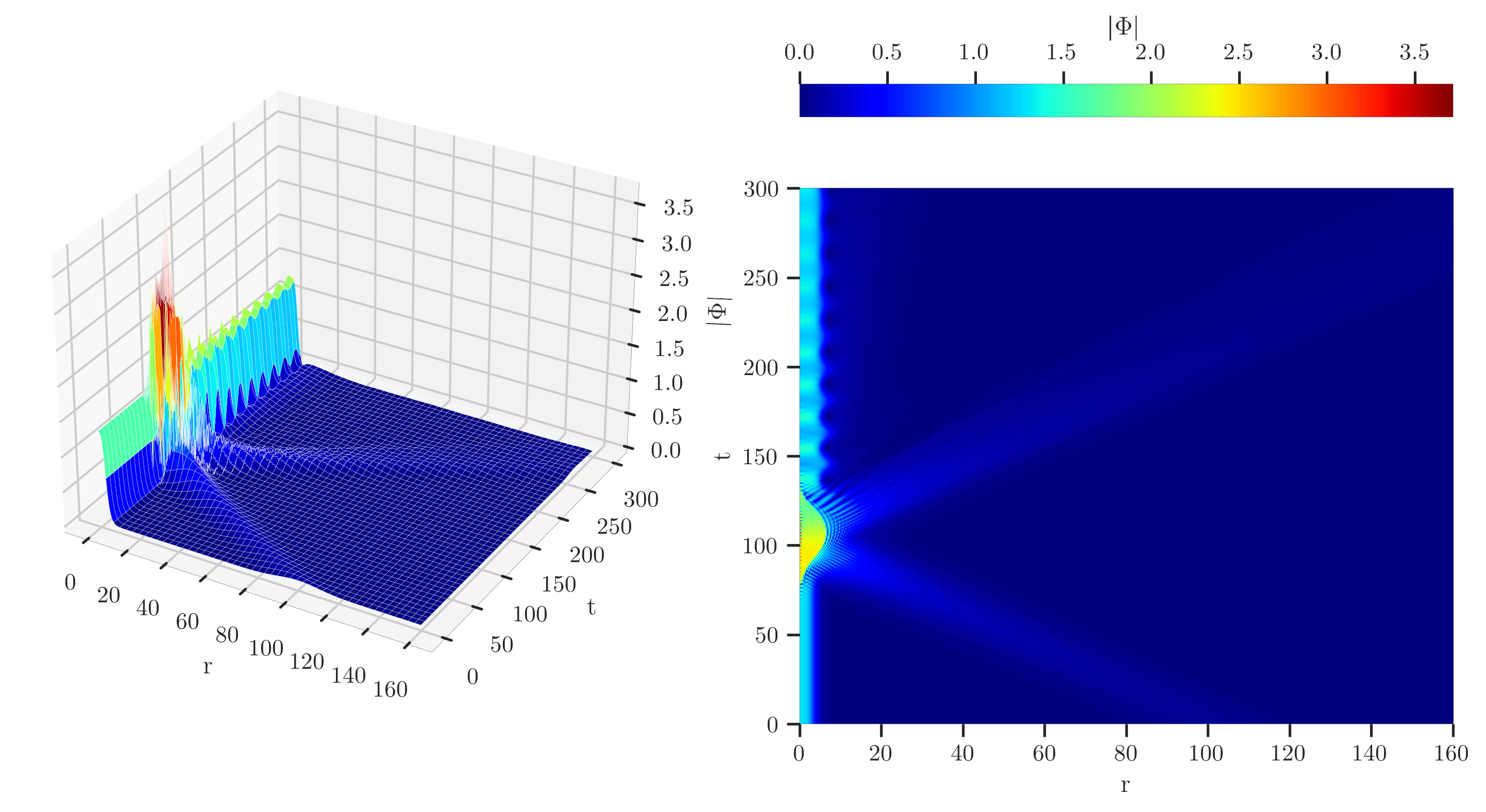}
	\caption[The solution of Q-ball scattering with spherical waves in spacetime lattice.]{The 3D plot and its corresponding heatmap to visualize the solution $\vert \Phi \vert$ of a Q-ball scattering with spherical waves.
Here, we artificially put large wave packet to visualize the propagation of the incoming and outgoing waves.
The scattering is very non-linear, which significantly deforms the Q-ball and makes it oscillate after the collision.
}
\end{figure}
In what follows, we pick the following value of the coupling $g=1/3$, and study only the spherically symmetric perturbations. 
During the lattice computations we will calculate the quantities $Z_E,Z_Q$ (see Eq. \ref{eq:1fieldZEZQ}) and compare them to the predictions of the linear analysis.
The energy current $J_E^r$ and the corresponding flux integrated over the time are given by: \begin{align}
	&J_E^r(r, t) = -2\,\mathrm{Re} (\partial_{r} \Phi^\ast \partial_{t} \Phi)
	= -2 \left( \partial_r \Phi_R \partial_t \Phi_R + \partial_r \Phi_I \partial_t \Phi_I \right), \\
	&E^{\mathrm{flux}}(r) = \int_0^t dt \oint d \bm{\Sigma} \cdot \bm{J}_E(\bm{x}, t)
	= A(r) \int_0^t dt \ J_E^r(r, t),
\end{align}
where we employed the spherical symmetry to factor out the area of the sphere $A(r)$.
Similarly for the flux of $U(1)$ global charge
\begin{align}
	&J_Q^r(r, t) = - 2\, \mathrm{Im}(\Phi^\ast \partial_t \Phi) 
  = -2 \left( \Phi_R \partial_t \Phi_I - \Phi_I \partial_t \Phi_R \right), \\
	&Q^{\mathrm{flux}}(r) = \int_0^t dt \oint d \bm{\Sigma} \cdot \bm{J}_Q (\bm{x}, t)
	= A(r) \int_0^t dt \ J_Q^r(r, t).
\end{align}
Thus the energy and charge amplifications are given by
\begin{align}
	Z_E = -\dfrac{E^{\mathrm{flux}}(t=\infty,R)}{E^{\mathrm{flux}}(t=t_*,R)}, ~~~~Z_Q = -\dfrac{Q^{\mathrm{flux}}(t=\infty,R)}{Q^{\mathrm{flux}}(t=t_*,R)}. 
\end{align}
We take  $t_* \simeq r_0-R +\mathcal{O}(10)\sigma_r$ in order to make sure that  we measure the flux after the initial wavepacket has passed through the radius $R$.
In our analysis we will focus primarily on the validity of the linear perturbation discussion and we report it as a function of the perturbation size.
Since the amplitude of the perturbation will grow as $1/r$ as the wavepacket approaches the Q-ball, it is better to use a rescaled version of it 
\bea
\delta_r\equiv \frac{\delta \times r_0}{R_Q\times \phi_Q(0)},
\eea
where $R_Q$ is the radius when the Q-ball field becomes equal to the half of its value in the center.
The results are shown on the Fig.~\ref{fig: Zratio delta}, where we indicate the discrepancies between the linear and non-linear approaches.
\begin{figure}[htbp]
	\begin{subfigure}{\textwidth}
		\centering
		\includegraphics[width=\linewidth]{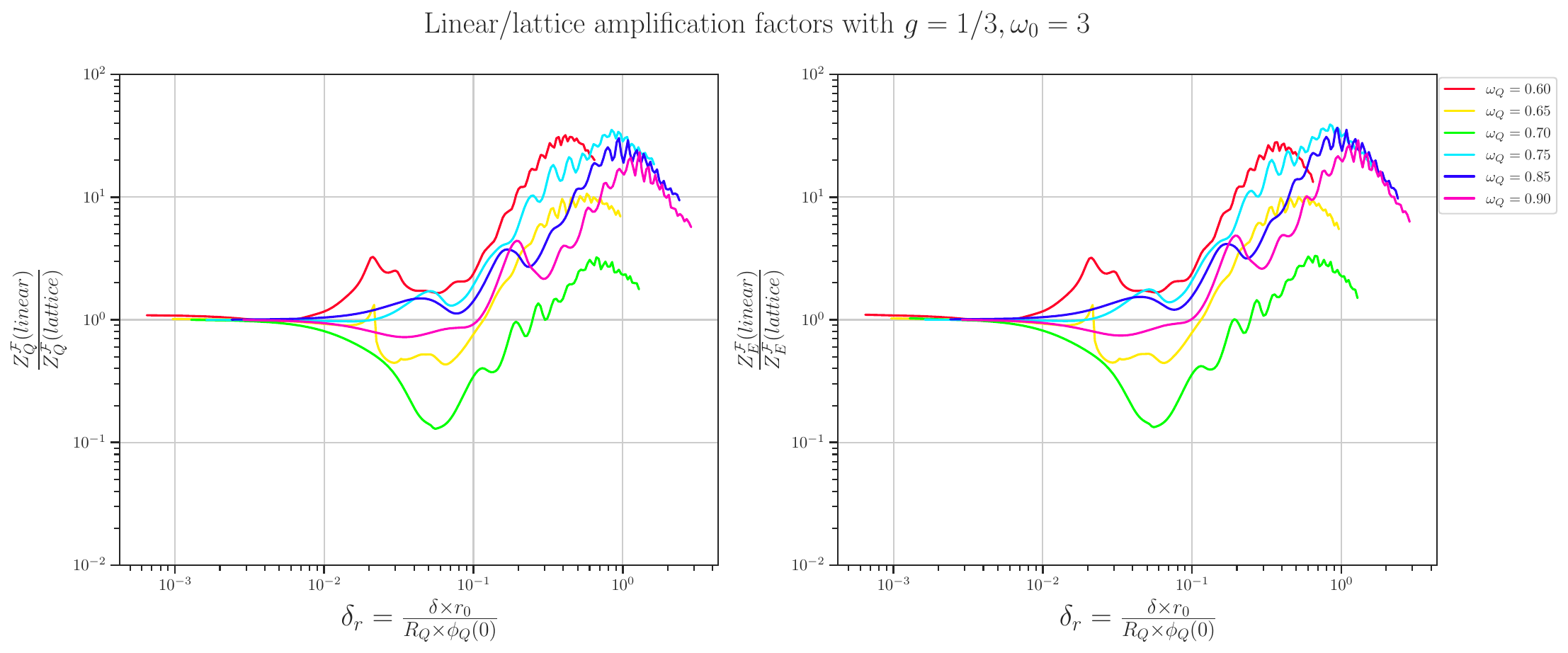}
	\end{subfigure}\\[10pt]
    \begin{subfigure}{\textwidth}
		\centering
\includegraphics[width=\linewidth]{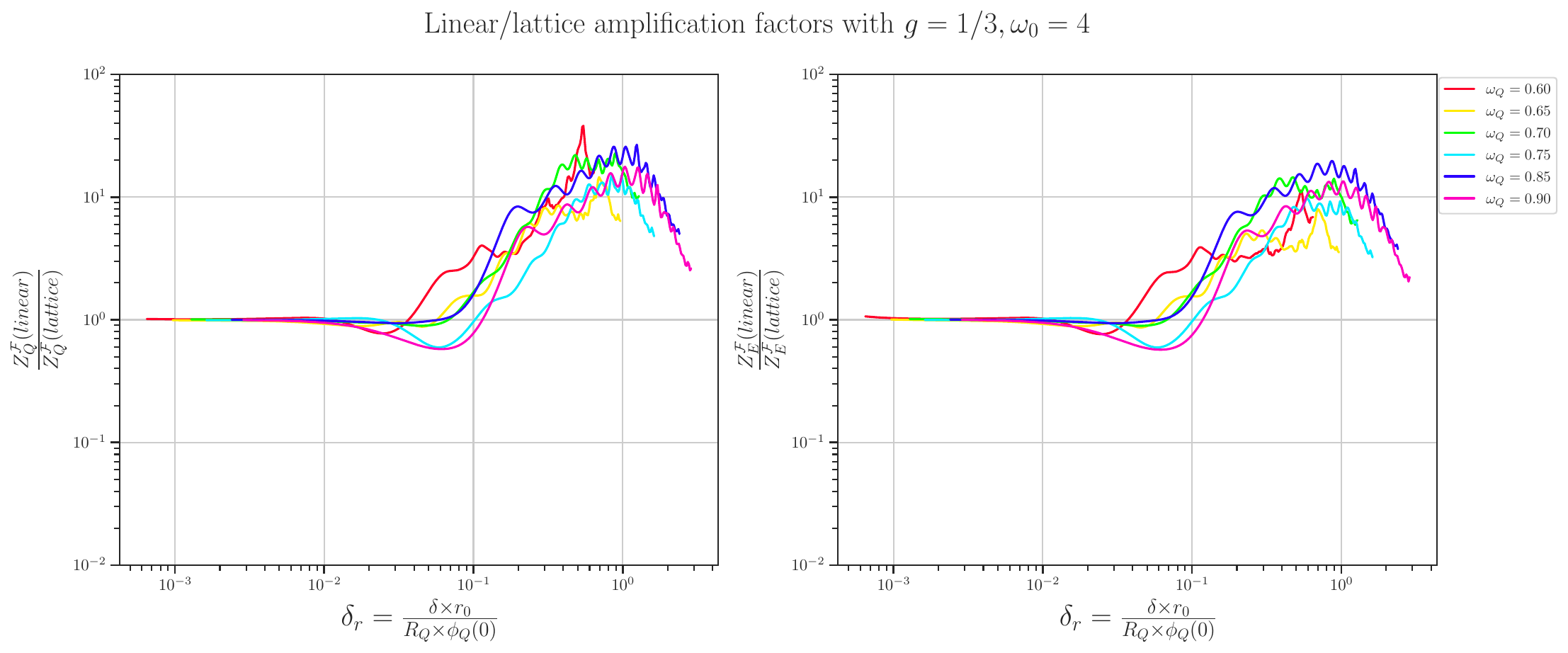}
	\end{subfigure}
	\caption{Comparison of amplification factors between the linear regime and full lattice simulation results. 
	Here $R_Q$ is defined as a radius where the field becomes half of the value at the center.
	Recall that the height of the Q-ball profile with given parameters is $\mathcal{O}(1)$, hence $\delta_r$ can also be thought of as the relative height between the incoming wavepacket and the Q-ball background.
	We notice that the convergence rate towards small $\delta_r$ is slightly faster with $\omega_0 = 4$ compared to $\omega_0 = 3$ due to the threshold effects.
}
  \label{fig: Zratio delta}
\end{figure}
These effects, as expected,  grow with the larger size of the perturbations. Additionally there are some sub-leading technical  effects feeding the discrepancies between linear and lattice analysis, which we list below:

\begin{itemize}
	\item The wave-packet width is not sufficiently large and hence we lose the resolution of the incoming wave momenta:
This effect becomes particularly important if $\omega$ is close to the threshold of having two propagating modes $\pm$, since one of the $k_\pm$ becomes very small and we need to make sure having broad enough wavepacket $\sigma_r$ in order to capture enough oscillations in space.
For example, in Fig.~\ref{fig: Zratio delta}, with $\omega_0=4$ we only need to put $\sigma_r=10$ and still obtain good agreement of linear/lattice.
Meanwhile, with $\omega_0=3$, we are closer to the limit $\omega = 1 + \omega_Q$ (recall that in our set-up, $\omega_0 = \omega_+ = \omega_Q + \omega$). 
In order to obtain somewhat comparable precision with $\omega_0 \geq 4$, we need to tune $\sigma_r = 30$, and consequently increase the lattice size to correctly capture the whole incoming-outgoing wavepackets.
	\item Precision of initial Q-ball profile: the profile of the initial Q-ball must be exact, otherwise it will radiate energy and charge and this becomes a background for the ``measurement" of the charge and energy extractions during the interaction with the perturbations.
	\item Discretisation of time and space leading to the generic systematic error of lattice results. 
\end{itemize}

\subsection{Physics implications}
We find that the linear analysis  results remain valid up to the perturbations of order $\delta_r\lesssim 10^{-2}$. 
This leads to the natural question of whether such a regime is realized in the early universe.
In order to effectively describe particle interactions with Q-balls, we must set the charge to be equal to one.
Consequently, the constraint on the amplitude becomes a constraint over the wave packet size (coherence length) $L_\omega$, indeed
\bea
Q=1\sim \frac{\mu^3}{\lambda}\delta_{\rm min}^2\phi_{Q}(0)^2 L_\omega ^3 \omega \Rightarrow L_\omega \gtrsim \l(\frac{\mu^3}{\lambda}\delta_{\rm min}^2 \phi^2_{Q}(0)\omega\r)^{-1/3},
\eea
where $\phi_Q(0)$ and $\omega$ are rescaled dimensionless (see Eq.\ref{eq:ac-dimless}) . Assuming the coherence length is of the order of the the inverse plasma temperature (we expect the wave packet length to be roughly of order of mean free path) we arrive at the condition 
\bea
\frac{\lambda^{1/2}T^{3/2}}{\mu^{3/2}}\times \l[ \phi_Q(0)\omega^{1/2}\r]^{-1}< 10^{-2}.
\eea
So once the temperature drops by one order of magnitude below the typical energy scale of the Q-ball field  ($\mu$) we expect the linear approximation to lead to the reliable results. 
For such temperatures the particle corresponding to the Q-ball perturbation
is non-relativistic (unless we have have a very small coupling $\lambda$), then from Fig.~\ref{Fig:reflection} we expect that most of the particle Q-ball interactions will be either elastic (reflection) or leading to the Q-ball charge reduction.

\section{Two-field Q-balls}
\label{sec:2field}
\begin{figure}[H]
	\centering
	\includegraphics[scale=0.6]{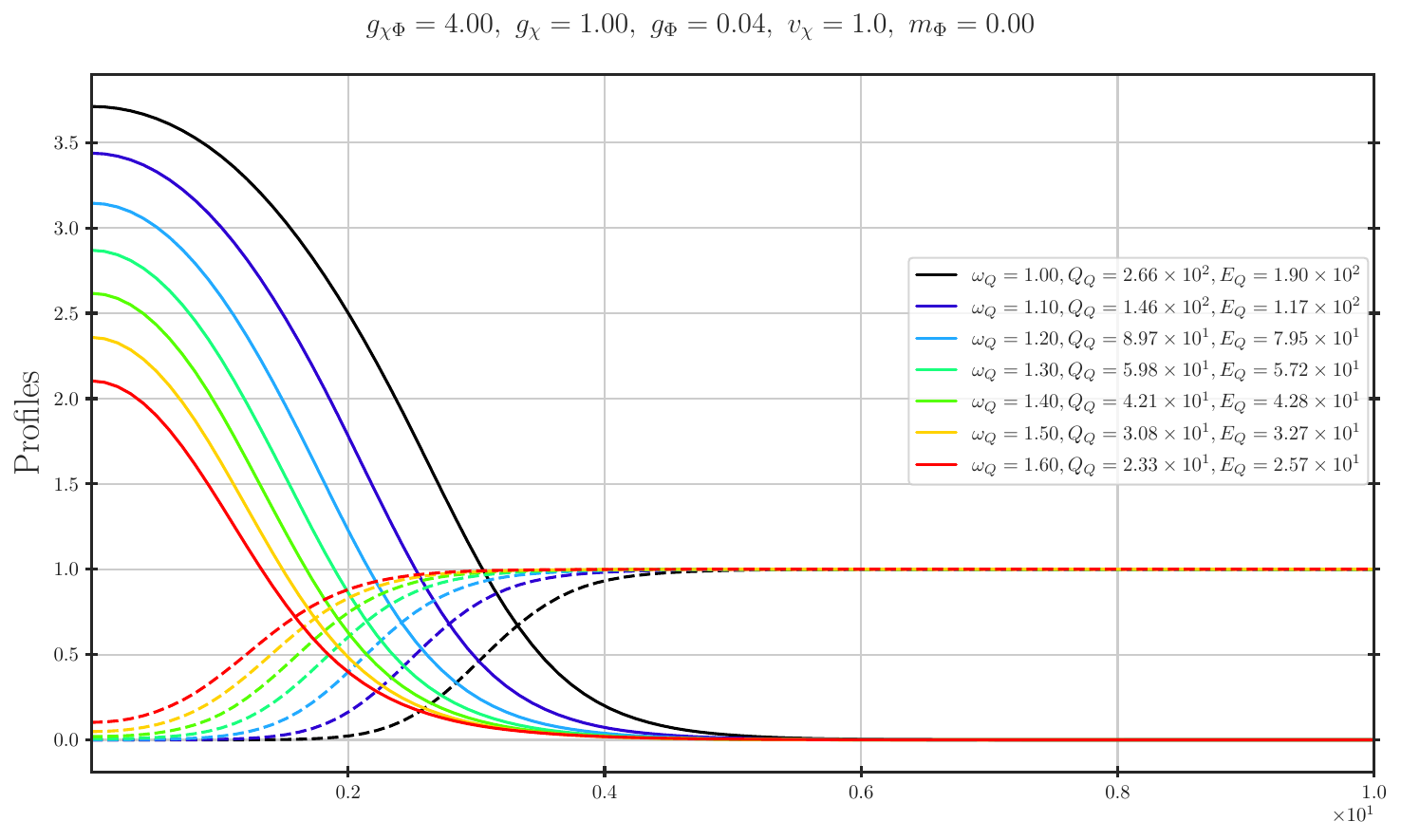}
	\caption{The profiles of FLS solitons ($\phi_Q$: solid line, $\chi_Q$: dashed line) with various internal frequencies $\omega_Q$, with reported total charge $Q_Q$ and total energy $E_Q$. All these profiles pass the sanity check of stability under decaying into free particles, i.e $E_Q < Q_Q \sqrt{m_\Phi^2+g_{\chi\Phi} v_\chi^2}$.}
	\label{fig:qball-FLS}
\end{figure}

We proceed by extending the analysis of the previous section to the two-field Q-ball case. 
This type of models where first studied by Friedberg, Lee and Sirlin (FLS)\cite{Friedberg:1976me,Friedberg:1976az,Friedberg:1976ay} (see for review \cite{Lee:1991ax}). 
Interestingly, this class of models does not require non-renormalizable interactions for the Q-ball solutions, and instead the non-linearities required for soliton formation arise from the non-linear couplings among the various field components.
These can be easily realized in the BSM scenarios, and for a long time has attracted the attention of model builder for DM prospective \cite{Bai:2022kxq,Ponton:2019hux,Krylov:2013qe,Jiang:2024zrb,Lennon:2021zzx,Bishara:2021fag,Bai:2021xyf,Bai:2019ogh}.
The simplest realization consists of the complex field $\Phi$ coupled to a real scalar $\chi$ fields:
\begin{align}
	\mathcal{L} = |\partial_\mu\Phi|^2 + \frac{1}{2}\,(\partial_\mu\chi)^2 - V (|\Phi|,\chi).
\end{align}
The potential breaks spontaneously the $\mathbb{Z}_2$ symmetry $\chi\to-\chi$ and generically can be written as follows
\begin{align}
	V(\abs{\Phi}, \chi) &= g_{\chi \Phi} \chi^2 \abs{\Phi}^2 + g_\chi \left( \chi^2 - v_\chi^2 \right)^2 + m_\Phi^2 \abs{\Phi}^2 + g_\Phi \abs{\Phi}^4.
\end{align}
The corresponding equations of motion are:
\begin{align}
	\begin{cases}
		\square \chi + \dfrac{\partial V}{\partial \chi} = 0 \\[10pt]
		\square \Phi + \dfrac{\partial V}{\partial \Phi^\ast} = 0
	\end{cases}
	\quad \Rightarrow \quad
	\begin{cases}
		\left[ \square + 2 g_{\chi \Phi} \abs{\Phi}^2 + 4 g_{\chi} (\chi^2 - v_\chi^2) \right] \chi = 0 \\[10pt]
		\left[ \square + g_{\chi\Phi} \chi^2 + m_\Phi^2 + 2 g_\Phi \abs{\Phi}^2 \right] \Phi = 0
	\end{cases}. 
\end{align}
This system admits a spherically symmetric Q-ball solution 
\begin{align}
\label{eq:2field-qball}
	\Phi_Q(\bm{x}, t) = \dfrac{\phi_Q(r)}{\sqrt{2}} e^{-i \omega_Q t}, \qquad
	\chi_Q(\bm{x}, t) = \chi_Q(r),
\end{align}
with $\phi_Q(r) \in \mathds{R}$, 
where at infinity $\chi_Q \rightarrow v_\chi, \phi_Q \rightarrow 0$  at $r\to \infty$ (see Fig.~\ref{fig:qball-FLS} for typical solutions,which  we have found  using the path deformation method).

\subsection{Linear perturbation}
\begin{figure}[H]
	\begin{subfigure}{\textwidth}
		\centering
\includegraphics[width=\textwidth]{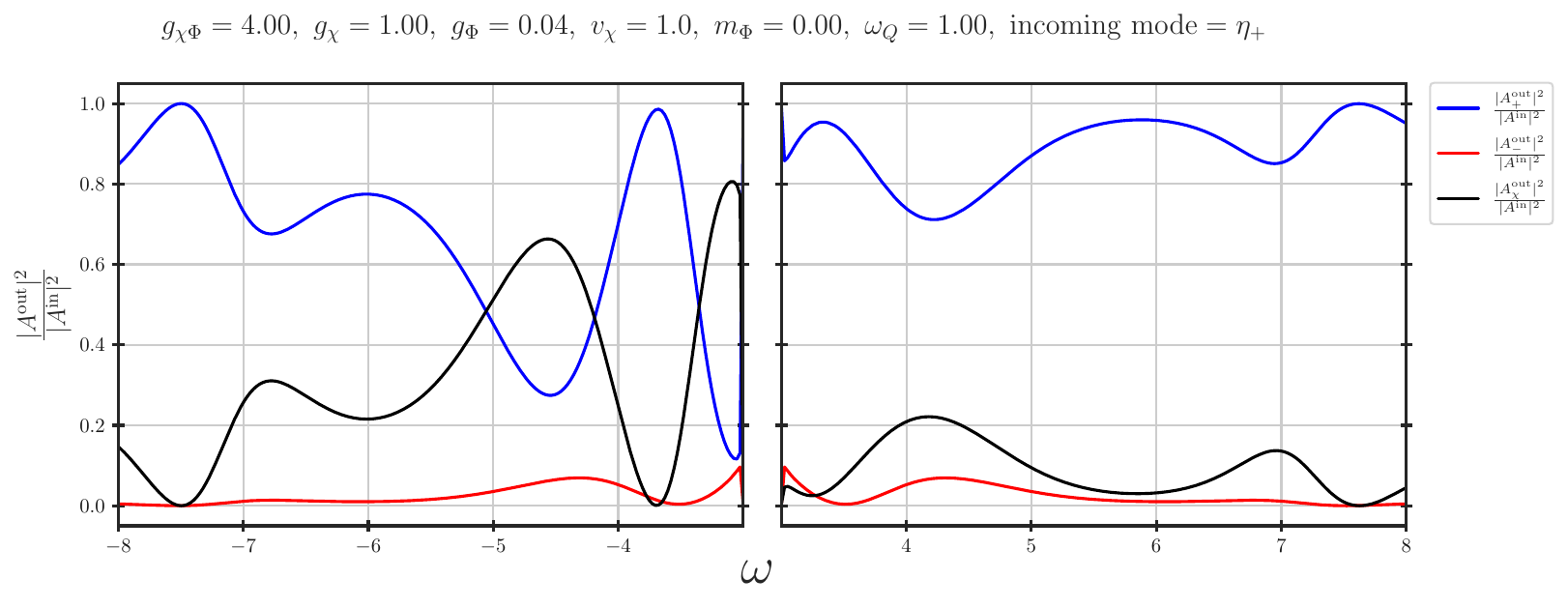}
	\end{subfigure}
	\begin{subfigure}{\textwidth}
		\centering
\includegraphics[width=\textwidth]{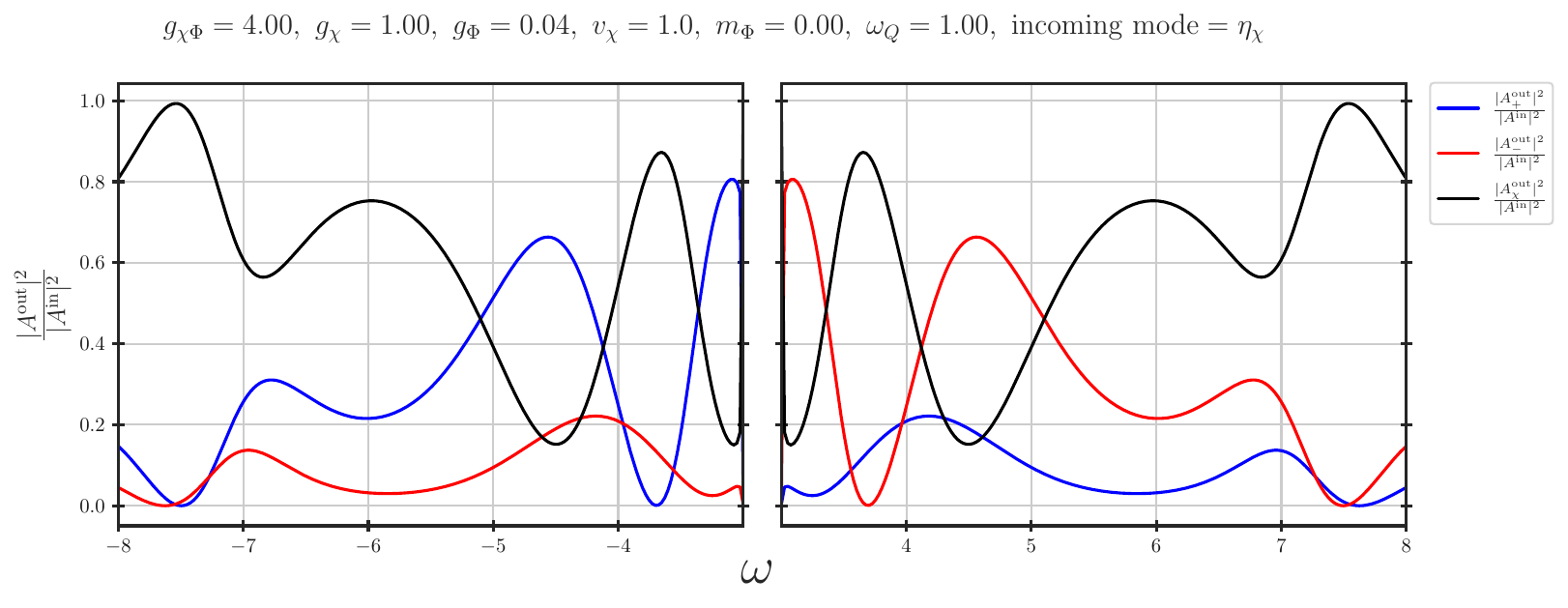}
	\end{subfigure}
	\caption{Ratios of the expansion coefficents, with incoming mode $\eta_+$ (top plot) and $\eta_\chi$ (bottom plot).
}
	\label{fig: Aratio FLS model}
\end{figure}

\begin{figure}[H]
	\centering
	\includegraphics[width={\textwidth}]{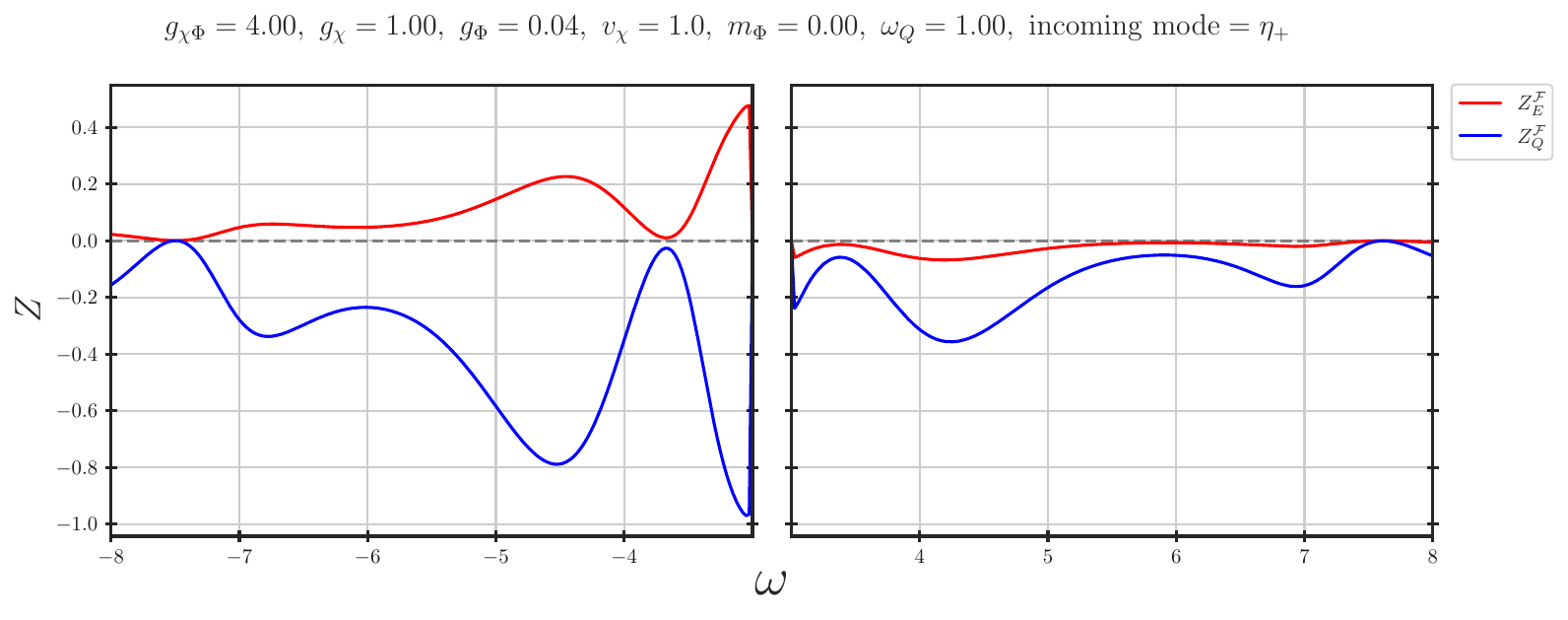}
	\caption{Energy and charge amplification factors with incoming mode $\eta_+$.}
	\label{fig: FLS Z incoming plus}
\end{figure}

\begin{figure}[H]
	\centering
	\includegraphics[width={\textwidth}]{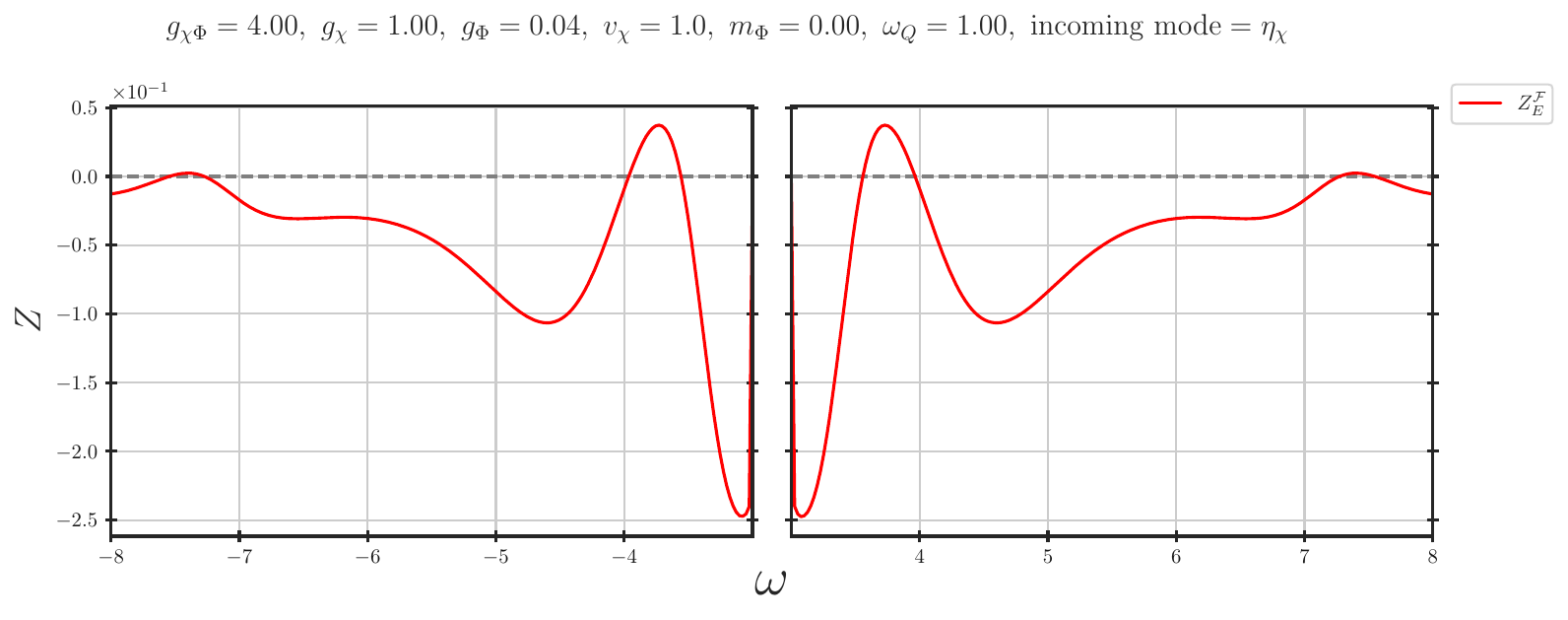}
	\caption{Energy amplification factors with incoming mode $\eta_\chi$.}
	\label{fig: FLS Z incoming chi}
\end{figure}

Once the solution for the Q-balls are obtained, we can proceed to the analysis of the perturbations. 
This analysis follows the discussion in the Section.~\ref{sec:1field}, and the only difference is the perturbation of the real field, which we need to take into account.
Denoting the perturbations as $\Phi = \Phi_Q + \Phi_1$ and $\chi = \chi_Q + \chi_1$, 
the following equations of motion are obtained:
\begin{align}
	&\begin{cases}
		\left[ \square +\dfrac{\partial^2 V}{\partial \chi^2} \biggr\vert_{(\chi, \Phi) \rightarrow (\chi_Q, \Phi_Q)} \right] \chi_1
		+\left[ \dfrac{\partial^2 V}{\partial \chi \partial \Phi} \biggr\vert_{(\chi, \Phi) \rightarrow (\chi_Q, \Phi_Q)} \Phi_1 + h.c \right]
		= 0 \\[15pt]
		\left[ \square + \dfrac{\partial^2 V}{\partial \Phi^\ast \partial \Phi} \biggr\vert_{(\chi, \Phi) \rightarrow (\chi_Q, \Phi_Q)} \right] \Phi_1
		+\dfrac{\partial^2 V}{\partial (\Phi^\ast)^2} \biggr\vert_{(\chi, \Phi) \rightarrow (\chi_Q, \Phi_Q)} \Phi^\ast_1 
		+ \dfrac{\partial^2 V}{\partial \Phi^\ast \partial \chi} \biggr\vert_{(\chi, \Phi) \rightarrow (\chi_Q, \Phi_Q)} \chi_1
		= 0 .
	\end{cases} 
\end{align}
Using the explicit form for the potential and the Q-ball solution Eq.~\eqref{eq:2field-qball} we get:
\begin{align} 
	&\begin{cases}
		\left[ \square + g_{\chi \Phi} \phi_Q^2(r) + 4 g_{\chi} \left( 3 \chi_Q^2(r) - v_\chi^2 \right) \right] \chi_1
		+ \sqrt{2} g_{\chi \Phi} \chi_Q(r) \phi_Q(r) \left( \Phi_1 e^{+i \omega_Q t} + \Phi^\ast_1 e^{-i \omega_Q t} \right) = 0 \\[5pt]
		\left[ \square + g_{\chi \Phi} \chi_Q^2(r) + m_\Phi^2 + 2 g_\Phi \phi_Q^2(r) \right] \Phi_1 + g_\Phi \phi_Q^2(r) e^{-2i \omega_Q t} \Phi_1^\ast + \sqrt{2} g_{\chi \Phi} \chi_Q(r) \phi_Q(r) e^{-i \omega_Q t} \chi_1 = 0
	\end{cases}. \label{eq: linearized eqn for FLS model}
\end{align}
This system admits time dependent  solutions of the following form:
\begin{align}
	&\chi_1(\bm{x}, t) = \eta_\chi(\bm{x}) e^{-i \omega t} + \eta^\ast_\chi(\bm{x}) e^{+i \omega t}, \qquad
	\Phi_1(\bm{x}, t) = \eta_+(\bm{x}) e^{-i \omega_+ t} + \eta_-(\bm{x}) e^{-i \omega_- t},
\end{align}
with $\omega_\pm \equiv \omega_Q \pm \omega$. 
We can see that comparing to the one complex field Q-ball discussion in the Section.~\ref{sec:1field}, here the modes with three different energies are mixed.
By substituting this ansatz to \eqref{eq: linearized eqn for FLS model}, we obtain the equations for different modes
\bea
\l\{\baa{c}
\left[ \nabla^2 + \omega^2 - g_{\chi \Phi} \phi_Q^2 - 4 g_{\chi} \left( 3 \chi_Q^2 - v_\chi^2 \right) \right] \eta_\chi - \sqrt{2} g_{\chi \Phi} \chi_Q \phi_Q \left( \eta_+ + \eta^\ast_- \right) = 0  \\
	\left[ \nabla^2 + \omega_+^2 - g_{\chi \Phi} \chi_Q^2 - m_\Phi^2 - 2 g_\Phi \phi_Q^2 \right] \eta_+ 
	- g_\Phi \phi_Q^2 \eta^\ast_-
	- \sqrt{2} g_{\chi \Phi} \chi_Q \phi_Q \eta_\chi = 0 \\
	\left[ \nabla^2 + \omega_-^2 - g_{\chi \Phi} \chi_Q^2 - m_\Phi^2 - 2 g_\Phi \phi_Q^2 \right] \eta_- 
	- g_\Phi \phi_Q^2 \eta^\ast_+
	- \sqrt{2} g_{\chi \Phi} \chi_Q \phi_Q \eta^\ast_\chi = 0.
    \eaa \r.
\label{eq:FLS-pert}
\eea
Similar to the one field Q-ball case at the linear perturbation analysis there is a conserved current, corresponding to the total particle number conservation and is given by:
\begin{align}
	\bm{J}_\eta = 2\,\text{Im}\l( \eta_\chi \bm{\nabla} \eta_\chi^\ast +\eta_+\bm{\nabla} \eta_+^\ast-\eta_-\bm{\nabla} \eta_-^\ast \r), \qquad \bm{\nabla} \cdot \bm{J}_\eta=0
. \label{eq: particle number flux for eta FLS}
\end{align}
Far from the Q-ball the mixing between various modes is switched off and the functions $\eta$ have the following form: 
\bea
\label{eq:perturbations-momentum}
&&\begin{cases}
\eta_\chi(r \rightarrow \infty)
		= \dfrac{1}{\sqrt{\abs{k_\chi}}\,r}  \left( A_\chi^{\rm out} e^{i k_\chi r}
		+ A_\chi^{\rm in} e^{-i k_\chi r} \right) \\[10pt]
\eta_\pm(r \rightarrow \infty)
		= \dfrac{1}{\sqrt{\abs{k_\pm}}\,r } \left( A_\pm^{\rm out} e^{i k_\pm r}
		+ A_\pm^{\rm in} e^{-i k_\pm r} \right)
		\end{cases}\nn
&&	k_\chi \equiv s_\omega \sqrt{\omega^2 - 8 g_{\chi} v_\chi^2}, \qquad
k_\pm \equiv \pm s_\omega \sqrt{\omega_\pm^2 - g_{\chi \Phi} v_\chi^2 - m_\Phi^2}.
\eea
$\bm{J}_\eta$ current conservation in this case leads to the relation:
	\begin{align}
		\abs{A_\chi^{\rm out}}^2 + \abs{A_+^{\rm out}}^2 + \abs{A_-^{\rm out}}^2
		= \abs{A_\chi^{\rm in}}^2 + \abs{A_+^{\rm in}}^2 + \abs{A_-^{\rm in}}^2. \label{eq:conservation of particle number FLS}
	\end{align}
Similarly to the one field case we can define $S$ matrix relating the 
\bea
\vec A^{\, \rm out}=S \vec A^{\, \rm in},~~A^{\rm out, in}\equiv \l[\baa{c}
A_+^{\rm out, in}\\
(A_-^{\rm out, in})^*
\\
A_\chi^{\rm out, in}
\eaa\r].
\eea
The S-matrix is symmetric and unitary, which leads to the following relations between the $A^{\rm in, out}$
	\begin{align}
		\abs{\dfrac{A^{\rm out}_+}{A^{\rm in}_-}}^2_{A^{\rm in}_{+,\chi}=0} = \abs{\dfrac{A^{\rm out}_-}{A^{\rm in}_+}}^2_{A^{\rm in}_{\chi,-}=0}, 
		\abs{\dfrac{A^{\rm out}_+}{A^{\rm in}_\chi}}^2_{A^{\rm in}_{+,-}=0} = \abs{\dfrac{A^{\rm out}_\chi}{A^{\rm in}_+}}^2_{A^{\rm in}_{\chi,-}=0}, 
		\abs{\dfrac{A^{\rm out}_-}{A^{\rm in}_\chi}}^2_{A^{\rm in}_{+,-}=0} = \abs{\dfrac{A^{\rm out}_\chi}{A^{\rm in}_-}}^2_{A^{\rm in}_{\chi,+}=0}.
	\end{align}
Additionally under the transformation  $\omega \leftrightarrow -\omega$ the $(+,-)$ are flipped leaving the $\chi$ unchanged. These two properties combined lead to the relation
\begin{align}
		\abs{\dfrac{A^{\rm out}_-(-\omega)}{A^{\rm in}_+(-\omega)}}^2_{A_{-,\chi}^{\rm in}=0} = \abs{\dfrac{A^{\rm out}_-(\omega)}{A^{\rm in}_+(\omega)}}^2_{A_{-,\chi}^{\rm in}=0}.
	\end{align}
Let us look at the energy exchange between the Q-ball and its perturbations. We will set $\omega_Q>0 $ and will assume 
 that $\omega>0$ ($\omega<0$ leads to the exchange of $\pm$ modes), then the energies of the coupled perturbations satisfy:
\begin{align}
\label{eq:3field-energy}
	\omega_+ > \omega > \abs{\omega_-}\Rightarrow E_+> E_\chi > E_-.
\end{align}
Let us assume that initial state is $\eta_-$, then the only possible reactions are:
\begin{align}
	& Q+\phi^\dagger \to (Q-2)+\phi\nn 
	& Q+\phi^\dagger \to (Q-1)+\chi.
\end{align}
The relations in Eq.~\eqref{eq:3field-energy} in both cases ($\phi$ and $\chi$ final states) predict energy extraction, and
the total energy release will be given by:
\begin{align}
	&1+Z_E
	= \abs{\dfrac{\omega}{\omega_+} +\dfrac{\omega_Q}{\omega_+} (1+Z_Q)}
	= \abs{1 + \dfrac{\omega_Q}{\omega_+} Z_Q},\nn
	&\text{where: } 1+Z_Q = |A_+^{\rm out}|^2-|A_-^{\rm out}|^2, \quad
	A_+^{\rm in}=1,A^{\rm in}_-=0.
	\label{eq: FLS AQ with incoming plus mode} 
\end{align}
Following the same logic one can see that if  the initial state is $\eta_+$ both of the possible reactions
\bea
Q+\phi \to (Q+1) +\chi\nn
Q+\phi \to (Q+2)+\phi^\dagger
\eea
proceed with the energy absorption.
If the initial state is $\chi$, then the energy enhancement will be equal to:
\begin{align}
	1+Z_E=\mathcal{A}_E^{\mathcal{F}} = \abs{1 + \dfrac{\omega_Q}{\omega} \left[\abs{\dfrac{A_+^{\rm out}}{A_\chi^{\rm in}}}^2  - \abs{\dfrac{A_-^{\rm out}}{A_\chi^{\rm in}}}^2 \right]}.
	\end{align}
One can see that depending on which of $A_+^{\rm out}$ or $A_-^{\rm out}$ is larger there can be energy release or absorption by the Q-ball.
Using the particle language, there are two possible reactions 
\begin{align}
	&Q+\chi\to (Q+1)+\phi^\dagger \hbox {  energy absorption}, \nn
	&Q+\chi\to (Q-1)+\phi\hbox{  energy release,}
\end{align}
and depending on their rates, the energy will be either absorbed or released.	
The ratio of the incoming and outgoing expansion coefficients $A^{\rm in,out}_{\pm,\chi}$ can be calculated numerically and we report our results  for a particular set of the couplings on the Fig. \ref{fig: Aratio FLS model}.
Once  $A^{\rm in,out}_{\pm,\chi}$ are known, it is straightforward to compute the amplification factors for energy and charge, see numerical results on Fig.~\ref{fig: FLS Z incoming plus} and Fig.~\ref{fig: FLS Z incoming chi} for a specific scattering off Q-ball.
As expected from the previous discussion the energy is amplified with $\omega<0$ and attenuated with $\omega>0$, meanwhile the charge of incoming mode is always attenuated.
Typically, the amplification effects are stronger compared to the attenuation ones, and these energy/charge exchange mechanisms are much weaker for high frequency incoming mode.
At frequency $\omega$ very close to the threshold of having two propagating charged modes, both $Z_E$ and $Z_Q$ drops to zero. 
With frequencies lower than this threshold, we have elastic scattering where no energy and charge transfer occur.
On the Fig.~\ref{fig: FLS Z incoming chi} we present the energy extraction for the $\eta_{\chi}$ incoming mode. The symmetry of $Z$ under $\omega \to -\omega$ follows the symmetry properties of the $S$ matrix. Note that the sign of $Z$ is not fixed and we can have both energy absorption and extraction depending on the energy of the initial mode.
Since the incoming charge is $0$, it is not well-defined to talk about $Z_Q$ here.

\subsubsection{Propagating and bounded modes modes}
Depending on the parameters of the models some of the three modes ($\eta_\pm,\eta_\chi$), which can transform into each other will get bounded. Based on this we can indicate the various parameter space regions depending on the propagating fields content. Let us choose $\omega_Q>0$ then the conditions will be the following:
\bit
\item $\omega^2 > 8 g_\chi v_\chi^2~~\& ~~(\omega\pm\omega_Q)^2>g_{\chi\Phi} v_\chi^2+m_\Phi^2$:  ~all three modes ~$\eta_\pm,\eta_\chi$ are propagating.

\item $\omega^2 <8 g_\chi v_\chi^2~~\& ~~(\omega \pm\omega_Q)^2> g_{\chi\Phi} v_\chi^2+m_\Phi^2$:~~  $\eta_\pm$ are propagating, $\eta_\chi$ is bounded.

\item  $\omega^2 >8 g_\chi v_\chi^2~~\&  ~~(\omega\pm\omega_Q)^2 < g_{\chi\Phi} v_\chi^2+m_\Phi^2$: $\eta_\pm,\eta_\chi$ are propagating,$\eta_\mp$ is bounded

\item $\omega^2 <8 g_\chi v_\chi^2~~\&  ~~(\omega\mp\omega_Q)^2 < g_{\chi\Phi} v_\chi^2+m_\Phi^2 ~\&~
(\omega \pm\omega_Q)^2>{g_{\chi\Phi} v_\chi^2+m_\Phi^2}$:
 $\eta_\pm$ is propagating, $\eta_{\mp},\eta_\chi$ are bounded.

\item $\omega^2 >8 g_\chi v_\chi^2~\&~ 
(\omega\pm\omega_Q)^2 < g_{\chi\Phi} v_\chi^2+m_\Phi^2$:
only $\eta_\chi$ is propagating.
\item 
$\omega^2 <8 g_\chi v_\chi^2~\&~ 
(\omega \pm\omega_Q)^2< g_{\chi\Phi} v_\chi^2+m_\Phi^2$
all of the modes are bounded.
\eit
\begin{figure}[H]
	\centering
	\includegraphics[width={\textwidth}]{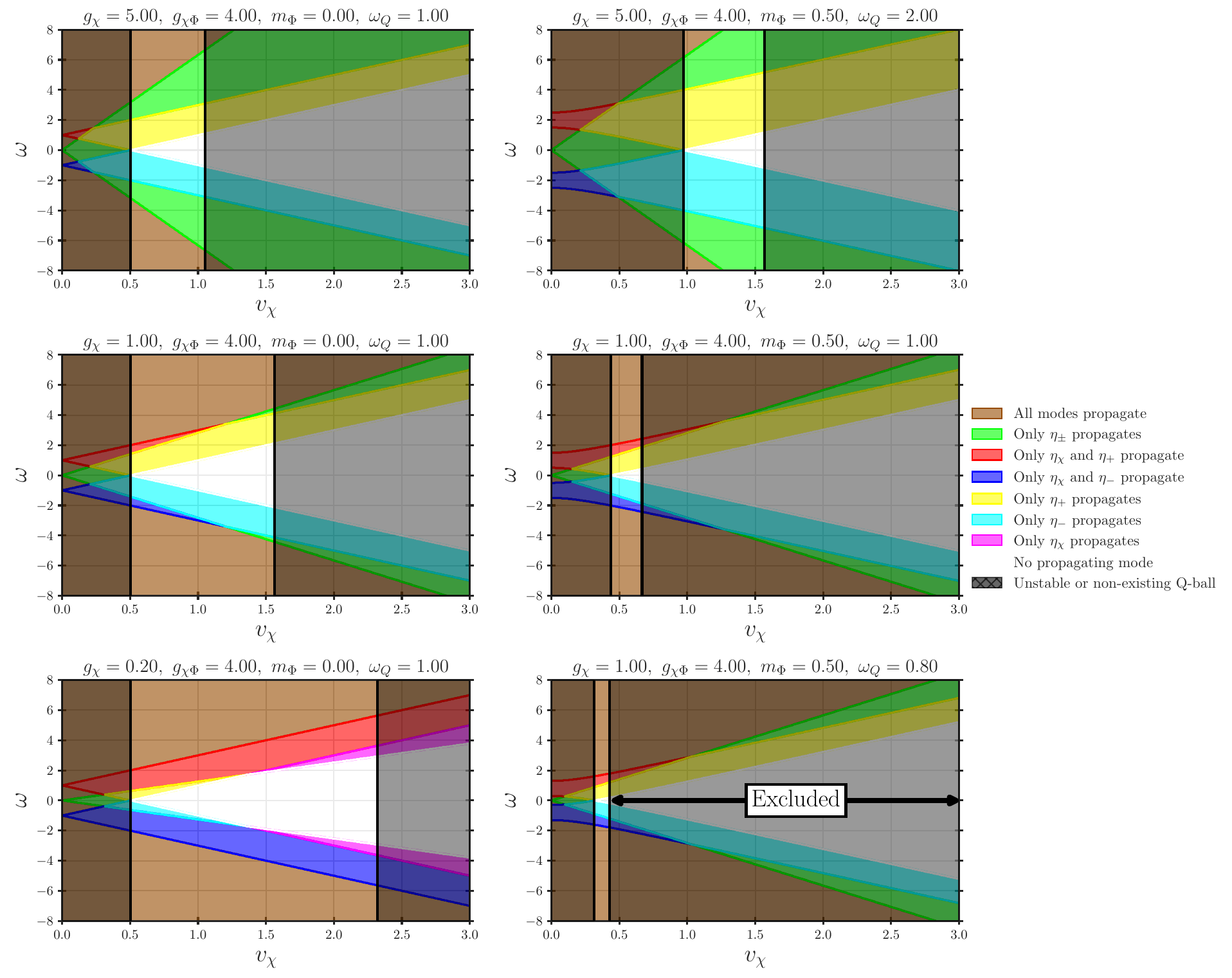}
	\caption{Constraints on $(v_\chi, \omega)$ on different propagating modes as a ``color mixer" plot. 
The first column corresponds to $m_\Phi = 0$ with varying $g_\chi$, while the second column corresponds to $m_\Phi = 0.5$ with varying $\omega_Q$. 
The gray shaded region indicates parameter space 
where a Q-ball profile is either non-existent, or  not stable under decay into quanta, for the  fixed $g_\phi$ coupling  value $g_\Phi = 0.04$ . Note that modification of this coupling only will alter the gray area,
while leaving everything else unchanged. 
The parameter set that we consistently used in the previous calculations corresponds to the middle left figure.
}
\label{fig: propagating modes constraints}
\end{figure}
These regions are visualized on the Fig.~\ref{fig: propagating modes constraints}, where the symmetry $(\eta_\pm, \eta_\chi) \rightarrow (\eta_\mp^\ast, \eta_\chi^\ast)$ under $\omega \rightarrow - \omega$ is reflected by the symmetricity of color patterns around the axis $\omega = 0$.
Furthermore, for arbitrary potential parameters, with sufficiently high $\omega$ one can always find that all modes propagates.
In opposite, with fixed $\omega$ and sufficiently large symmetry breaking scale $v_\chi$, we observe that all modes become bounded, providing that the Q-ball solutions exist for such potential.

\section{Summary}
\label{sec:summary}
In this paper, we expand upon the investigation of Q-ball perturbations, drawing on previous works \cite{Saffin:2022tub,Cardoso:2023dtm}, with a particular emphasis on non-rotating Q-balls in 3+1 dimensions.
Our findings enhance the understanding of Q-ball formation mechanisms in early universe cosmology, particularly within models involving solitosynthesis.

Compared to the original works, the novelty of our results lies in three distinct yet interconnected directions.
First, we  provide an intuitive explanation of the energy extraction process, linking it to differences in the binding energies of Q-balls with varying charges—analogous to energy release in nuclear reactions.
Second, we analyze in detail the physics of the linear perturbations. 
We have identified the symmetry properties of various transition rates and we have  carefully determined the validity range of linear perturbation analysis for Q-balls.
This enables us to estimate when, in the early universe, linear analysis leads to the reliable predictions for the Q-ball interactions with plasma particles. 
Interestingly we find that the scattering of the non-relativistic particles over the Q-ball most likely lead to its charge and energy reduction at the classical level.
Third, for the first time we have analyzed the perturbations of the FLS Q-ball solution. 
We have demonstrated that energy extraction occurs in this case via transformations among modes with three distinct energies.

\section*{Acknowledgements}
We would like to acknowledge support by the European Union - NextGenerationEU, in the framework of the PRIN Project “Charting unexplored avenues in Dark Matter" (20224JR28W)

\appendix

\section{Symmetry properties of scattering at linear order}
\label{app:symmetry}
The system of linear equations of motion that describe the scattering processes of quanta off a Q-balls is of the following form (see Eq.~\eqref{eq:1fieldpert},\eqref{eq:FLS-pert})
\begin{align}
\label{eq:pert-generic}
	\left[ \nabla^2 + O(r) \right] \vec{\eta}(r) = 0,
\end{align}
where  we have introduced the following notation:
$
\vec \eta\equiv (\eta_+,\eta_-^*,\eta_\chi)
$ and $O$ is some matrix which at large distances becomes diagonal 
\bea
\lim_{r\to \infty} O(r)\to K^2,~~
K\equiv \mathrm{diag}(k_1,\,k_2,\dots) 
\eea
where $k_i$ are the  momenta of the incoming and outgoing modes.
Since we are using $\eta_-^*$ (and not $\eta_-$) as a component of the $\vec \eta$ we can always choose all $k_i$ to be of the same sign, then at large distances the solutions are as follows:
\begin{align}
	&\vec{\eta}(r\to\infty) 
	= \dfrac{1}{\sqrt{{K}} r} e^{iKr} \vec{A}^{\rm \; out} +  \dfrac{1}{\sqrt{{K}} r} e^{-iKr}\vec{A}^{\rm \; in},\nn
&\vec{A}^{\rm in, out}\equiv (A_+,A_-^*,A_\chi)^{\rm in, out}.
\end{align}
Since the Eq.~\eqref{eq:pert-generic} is linear in perturbations, the incoming and outgoing amplitudes $\vec A$ will be related by some matrix $S$ (which is some analogue of the scattering  S-matrix).
\bea
\vec{A}^{\rm \; out} = S^{}\; \vec{A}^{\rm \; in}.
\label{eq: S-matrix}
\eea
This matrix in general will depend on the Q-ball potential and its charge and the frequency of the incoming mode, additionally it must satisfy the following properties:
\begin{itemize}	
	\item {The S-matrix is unitary}: this follows from the conservation of the $\bm J_\eta$ current
	\begin{align}
		\bm{J}_\eta = i \left[ \vec{\eta} \cdot \bm{\nabla}\vec{\eta}^{\,\dagger} - \vec{\eta}^{\,\dagger}\cdot \bm{\nabla} \vec{\eta} \right]
		= 2 \mathrm{Im} \left( \vec{\eta}^{\,\dagger} \cdot \bm{\nabla}\vec{\eta} \right).
	\end{align}
	By substituting the asymptotic solutions, we obtain 
	\begin{equation} 
		|\vec{A}^{\rm \, out}|^2=|\vec{A}^{\rm \, in}|^2
		\label{eq: particle number conservation for all l}
	\end{equation}
	This implies that
	\begin{equation} 
		S^{\,\dagger}\,S^{} - \mathds{1} = 0 \implies S^{-1} = S^{\,\dagger} 
	\end{equation}
	\item 
 \bm{The S-matrix is symmetric:}
This follows from the reality of function $O(r)$. Then we immediately see that if 
$\vec\eta(r)$ is a solution of the equations 
\eqref{eq:pert-generic} then $\vec\eta^*(r)$ will be a solution as well.
\begin{align}
	&\vec{\eta}^{\, \ast}(r\to\infty) 
	= \dfrac{1}{\sqrt{K} r} e^{-iKr} \vec{A}^{\rm \, out \ast} +  \dfrac{1}{\sqrt{{K}} r} e^{-iKr}\vec{A}^{\rm \, in \ast}\nn
   & \qquad \Rightarrow \qquad
	\vec{A}^{\rm \, in \ast} = S\, \vec{A}^{\rm \, out \ast}
\end{align}
where in the second line we used the fact that $S$ matrix connects 
amplitudes with negative and positive exponents. Then using the 
eqn.~\eqref{eq: S-matrix}, we obtain
	\begin{equation}
		S\,S^{\ast} = \mathds{1} \implies S^T = S^{}
	\end{equation}
	The unitarity and symmetricity of the S-matrix constrain it to have only $n (n+1)/2$ independent real parameters for the case where $\vec \eta$ is a $n$ dimensional vector.

	\item \bm{Transformation of S-matrix with $\mathds{Z}_2$}: for two-field model, under $\omega \rightarrow - \omega$ we have $(\eta_\pm, \eta_\chi) \rightarrow (\eta_\mp^*, {\eta}_\chi^*)$ up to a complex scaling factor and a complex conjugation. 
Such an observation comes trivially from the equations of motion Eq.~\eqref{eq:FLS-pert}.
This imposes the following constraints that are independent of the unitarity and symmetricity:
	\begin{align}
		&\abs{S^{++}(- \omega)} = \abs{S^{--}(\omega)}, \qquad
		\abs{S^{+ \chi}(- \omega)} = \abs{S^{- \chi}(\omega)}, \\
		&\abs{S^{+-}(- \omega)} = \abs{S^{+-}(\omega)}, \qquad
		\abs{S^{\chi \chi}(- \omega)} = \abs{S^{\chi \chi}(\omega)}.
	\end{align}
	It is straightforward to obtain the coresponding relation for the one-field case by restricting the indices of S-matrix to be either $+$ or $-$.
	
\item  \bm{S-matrix in limit of having bounded modes:} Let $\omega_j^{\rm bound}$ is the frequency where the corresponding wavenumber $k_j = 0$, i.e the mode $j$ becomes bounded. 
Assuming the solution of the bounded mode close to the Q-ball is still sufficiently small for the validity of linear perturbation; this solution drops exponentially with respect to $r$ and becomes negligible when we stay sufficiently far. Then, the particle number conservation \eqref{eq: particle number conservation for all l} implies there is no conversion between the propagating modes and the bounded modes.
Consequently we should obtain the following limits:
\begin{align}
	\abs{S^{fi}\left( \omega \rightarrow \omega^{\rm bound}_i \right)}
	= \abs{S^{fi}\left( \omega \rightarrow \omega^{\rm bound}_i \right)}
	= \delta_{fi}.
\end{align}
Especially in the one-field model with $\vec{\eta} = (\eta_+, \eta_-^\ast)$, $\omega_-^{\rm bound} = 1+\omega_Q$ and $\omega_+^{\rm bound} = -1-\omega_Q$, we have
\begin{align}
	\abs{S\left( \omega \rightarrow \omega_-^{\rm bound} \right)}
	= \abs{S\left( \omega \rightarrow \omega_+^{\rm bound} \right)}
	= \mathds{1}_{2 \times 2},
\end{align}
which have been verified numerically.
	\item \bm{S-matrix for high frequency scattering:} In the linearized equations, if $\abs{\omega}$ is sufficiently large, the matrix $O(r)$ is approximately diagonal for all $r$, meaning that all modes decoupled and we restore a $U(1)^n$ global symmetry with each $U(1)$ corresponds to a phase rotation of each mode.
We then have $n$ approximately conserved current for each mode, implying $\abs{A_i^{\rm out}} = \abs{A_i^{\rm in}}$ for $i=1, \dots, n$, and hence the S-matrix reduces to a diagonal matrix of phases, or equivalently
	\begin{align}
		\lim_{\omega \rightarrow \pm \infty} S(\omega) = \mathds{1}.
	\end{align}
	This fact has been observed numerically for both one-field and two-field models.
\end{itemize}

Using the S-matrix language, one can express the amplification factors of a spherical incoming wave in terms of the S-matrix elements
\begin{align}
	&\mathcal{A}_Q = \abs{\dfrac{\abs{\sum_i \alpha_i S^{+i}}^2 - \abs{\sum_i \alpha_i S^{-i}}^2}{\abs{\alpha_+}^2 - \abs{\alpha_-}^2}}, \\
	&\mathcal{A}_E = \abs{\dfrac{\omega_+ \abs{\sum_i \alpha_i S^{+i}}^2 - \omega_- \abs{\sum_i \alpha_i S^{-i}}^2 + \omega \abs{\sum_i \alpha_i S^{\chi i}}^2}{\omega_+ \abs{\alpha_+}^2 - \omega_- \abs{\alpha_-}^2 + \omega \abs{\alpha_\chi}^2}},
\end{align}
where $\alpha_i \equiv \dfrac{A_i^{\rm in}}{\sum_j A_j^{\rm in}}$ is the ``amplitude fraction" of the mode $i$, assuming $\sum_j A_j^{\rm in} \neq 0$.

In case of having exactly one monochromatic incoming wave $i$, we have $\alpha_j = \delta_{ji}$, and the expression above reduces to
\begin{align}
	&\mathcal{A}_Q = \abs{\abs{S^{+i}}^2 - \abs{S^{-i}}^2}, \qquad i \neq \chi, \\
	&\mathcal{A}_E = \abs{\dfrac{\omega_+}{\omega_i} \abs{S^{+i}}^2 - \dfrac{\omega_-}{\omega_i} \abs{S^{-i}}^2 + \dfrac{\omega}{\omega_i} \abs{S^{\chi i}}^2}, \qquad \forall i \in (\pm, \chi).
\end{align}

Interestingly for the one complex field case  the $S$ is a $2\times 2$ matrix  with only $3$ independent components. It can be fixed completely by solving the Eq.~\eqref{eq:1fieldpert} only for one set of the initial conditions at the origin $(r\to 0)$. 
Indeed in this case we will need to solve the system of equations
\bea
	\vec A^{\rm \, out} = S_{2\times 2} \vec A^{\rm \, in}
\eea
to find all of the entries of $S$. Naively it looks like there are four independent equations since $\vec{A}^{\rm in, out}$ are complex vectors. 
However one equation is redundant due to the unitarity of S-matrix, i.e.
\bea
\abs{\vec A^{\rm \, out}}^2=\abs{\vec A^{\rm \, in}}^2.
\eea
In the case of FLS model S-matrix is $3\times 3$ and has 6 independent real components. 
However solving Eq.~\eqref{eq:FLS-pert} with one initial condition leads only $(2n-1)=5$ equations which are not sufficient to fix the S-matrix.

\section{Scattering beyond S-wave}
\label{app:non-sphere}
In the main text we focused only on the S wave ($\ell=0$) scatterings. 
In this appendix we discuss briefly the case of non-spherical perturbations. 
We will focus only on the non-spinning Q-ball solutions (see Ref\cite{Zhang:2024ufh} for a recent study in a general case of the spinning Q-balls) and we find that results for $\ell\neq 0$ are qualitatively similar to the $\ell=0$ case.
Indeed performing the decomposition in terms of the angular momentum 
\bea
\eta_\pm (r,\theta,\varphi)=\sum_{\ell=0}^{\infty}\sum_{m=-\ell}^\ell \eta^{m \ell}_\pm(r)e^{i m\varphi}P_\ell^m(\cos\theta)
\eea
where $P_{\ell}^m$ are associated Legendre polynomials, we then obtain the following equation
\bea
\label{eq:ml-comp}
\l[\frac{d^2}{dr^2}+\frac{2}{r}\frac{d }{dr}-\frac{\ell (\ell+1)}{r^2}\r] \eta^{m \ell}_{\pm}+\l[\omega_\pm^2-U\r] \eta^{m \ell}_{\pm}-W  (\eta^{-m \ell}_{\mp})^\ast=0.
\eea
We can see  that there will be a similar current $\bm{J}_\eta^{\ell m}$ conservation for all the possible values of $\ell,m$. Asymptotic solutions to the $\eta^{m \ell}$ have exactly the same form as in Eq.~\eqref{eq:asympt-S}
\bea
\label{eq:asympt-lm}
\eta_{\pm}^{\ell m}= \frac{1}{\sqrt {|k_{\pm}|}\, r}\l[A_{\pm,(\ell, m)}^{\rm in} e^{-i k_\pm r} + A_{\pm,(\ell, m)}^{\rm out}e^{i k_\pm r}\r].
\eea
and the relation between the amplitudes becomes
\bea
\label{eq:number-conslm}
\boxed{
|A^{\rm out}_{+,(\ell,m)}|^2+|A^{\rm out}_{- (\ell,-m)}|^2=|A^{\rm in}_{+,(\ell,m)}|^2+|A^{\rm in}_{-(\ell,-m)}|^2.
}
\eea
We can see that Eq.~\eqref{eq:number-conslm} imposes the number conservation for the particles with given angular momentum.
Recently the Ref.~\cite{Zhang:2024ufh} have looked at the perturbations for the rotating Q-ball. 
In this case, states with different values of $\ell$ are mixed and the value $m$ of the particle of the $\eta_\pm$ fields are related by $m_\pm=m_Q\pm m$. 
However even in this case the total number of particles is conserved and the particle and antiparticle energies are fixed to be $ \omega_\pm$. 
At last note that we can define the $S$ matrix in this case as well
\bea
\vec{A}^{\, \rm out}_{\ell, m}= S_{\ell, m} \vec{A}^{\, \rm in}_{\ell, m},
\eea
and it will be symmetric and unitary as for the $\ell=0$ wave case.

\subsection{Total cross section}
\label{app:plane wave}
We can proceed to the analysis of the perturbations which can be described in terms of the plane waves. For example let us assume that incoming state is $\eta_+$,
 then the
solution of the  equations of motion for the perturbations  takes the form 
\bea
&&\eta_+(r,\theta) = \frac{1}{\sqrt{k_+}}\Big[{f_+}(\theta)\,\frac{e^{i k_+r}}{r} + e^{ik_+ z}\Big],\nn
 && \eta_-(r,\theta) = \frac{1}{\sqrt{|k_-|}}\,{f_-}(\theta)\, \frac{e^{i k_-r}}{r}.
    \label{plane-wave}
\eea
We expand the plane-wave and the scattering amplitudes in partial waves
\bea 
&& e^{ik_+z} = \sum_{\ell=0}^\infty (2\ell+1)\,i^\ell\,j_\ell(k_+r)\,P_\ell(\cos{\theta}), \nn
&& f_\pm(\theta) = \sum_{\ell=0}^\infty (2\ell+1)\,f_{\pm,\, \ell}\,P_\ell(\cos{\theta}).
\eea
Substituting this into ~\eqref{plane-wave}, and recall the asymptotic behavior of the spherical Bessel function $j_\ell(z) \sim \left( 2iz \right)^{-1} \left( i^{-\ell} e^{iz} - i^\ell e^{-iz} \right)$
\bea
&& \eta_+(r, \theta) = \frac{1}{\sqrt{k_+}}\sum_{\ell=0}^\infty (2\ell+1)\,\Bigg[\Big(f_{+,\,\ell}+\frac{1}{2i k_+}\Big)\,\frac{e^{ik_+r}}{r} + \frac{(-1)^{\ell+1}}{2i k_+}\,\frac{e^{-ik_+r}}{r} \Bigg]\,P_\ell(\cos{\theta}),\nn
&& \eta_-(r, \theta) =\frac{1}{\sqrt{|k_-|}}\sum_{\ell=0}^\infty (2\ell+1)\,f_{-,\,\ell}\,\frac{e^{ik_-r}}{r}\,P_\ell(\cos{\theta})
 \eea
Following the usual quantum mechanics procedure we arrive for the expression for  cross sections $\sigma_{+f}$ for producing $f=\pm$ states
\bea
\sigma_{++}=4\pi \sum_{\ell=0}^\infty (2\ell+1) \l|f_{+,\ell}\r|^2, \qquad
\sigma_{+-}=4\pi \sum_{\ell=0}^\infty (2\ell+1) \l|f_{-,\ell}\r|^2.
\eea
The coefficients $f_\ell$ can be expressed in terms of the  $A_{\pm,\ell}^{\rm in, out}$ introduced in the previous section as follows:
\bea
f_{-,\ell} =\frac{(-1)^{\ell+1}}{2 i k_+}\l.\frac{A_{- \ell}^{\rm out}}{A_{+, \ell}^{\rm in}}\r|_{A_{-, \ell}^{\rm in}=0}, \qquad
f_{+,\ell}=\frac{i}{2 k_+}\l[1+(-1)^\ell \frac{A_{+ \ell}^{\rm out}}{A_{+, \ell}^{\rm in}}\r]_{A_{-, \ell}^{\rm in}=0}.
\eea
One can similarly derive the cross sections with other incoming mode.
In terms of S-matrices, the cross sections $\sigma_{if}$ for various initial states $i$ and final states $f$ are

\begin{subequations}
	\begin{empheq}{align}
	&\sigma_{++}(E)=\frac{\pi}{(E^2-1)} \sum_{\ell = 0}^\infty (2\ell+1)
	\l|1+(-1)^\ell S_\ell^{++}(\omega=E-\omega_Q)\r|^2, \\
	&\sigma_{+-}(E)=\frac{\pi}{(E^2-1)} \sum_{\ell = 0}^\infty (2\ell+1)
	\l| S_\ell^{-+}(\omega=E-\omega_Q)\r|^2, \\
	&\sigma_{--}(E)=\frac{\pi}{(E^2-1)} \sum_{\ell = 0}^\infty (2\ell+1)
	\l|1+(-1)^\ell S_\ell^{--}(\omega=E+\omega_Q)\r|^2, \\
	&\sigma_{-+}(E)=\frac{\pi}{(E^2-1)} \sum_{\ell = 0}^\infty (2\ell+1)
	\l| S_\ell^{+-}(\omega=E+\omega_Q)\r|^2,
	\end{empheq}
\end{subequations}
with energy $E^2 = 1 + |\bm{k_\pm}|^2$ since we are using dimensionless action, see Eq.\eqref{eq:ac-dimless}.
Note that even though S- matrix is defined in terms of $(A_{-,l}^{\rm in,out})^*$ the equations above are still valid due to the absolute value squared.

\begin{figure}[H]
	\centering
	\includegraphics[scale=0.6]{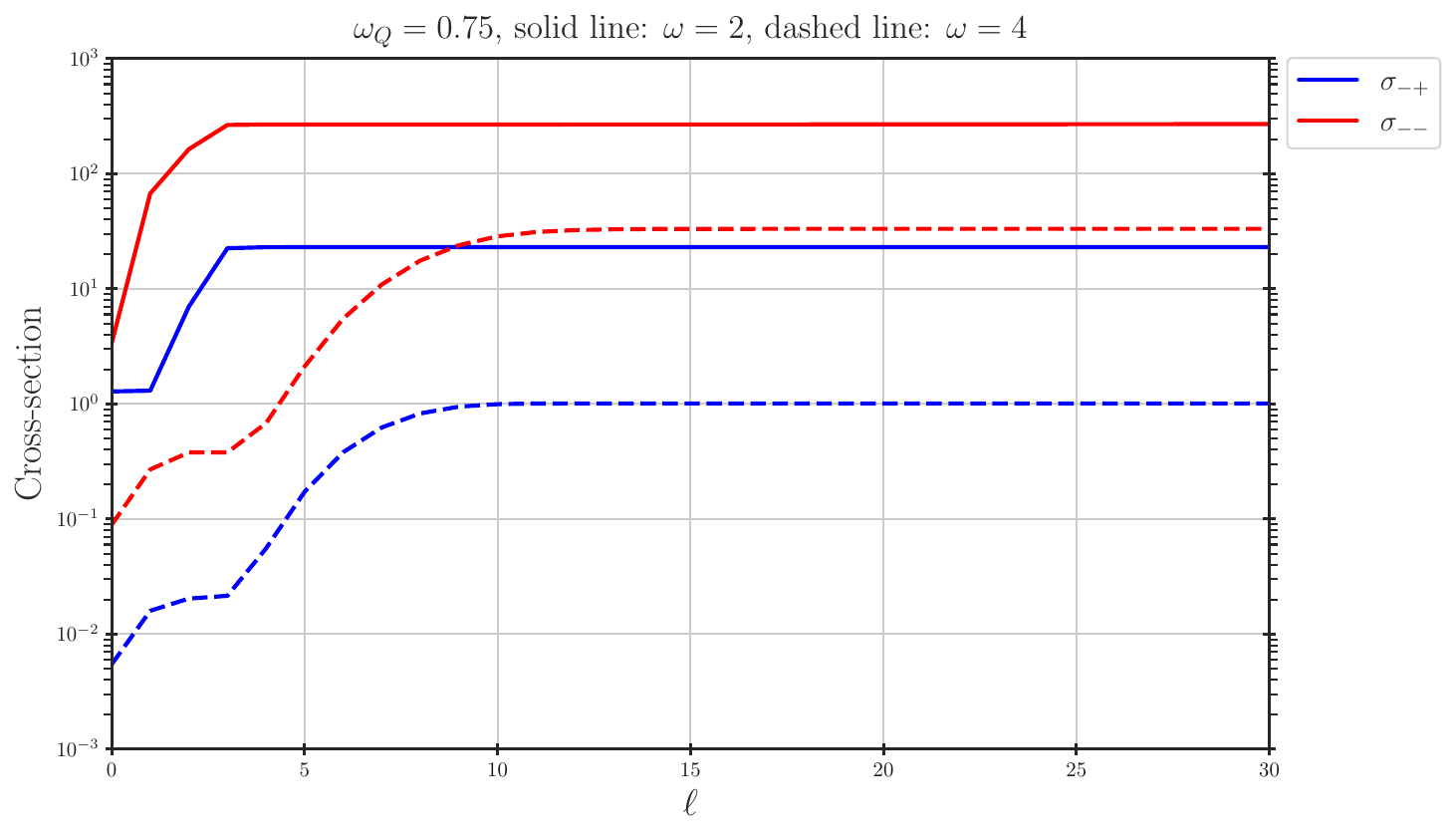}
 \caption{The saturation of the total cross section due to the scattering of an incoming particle ($-$ mode) against a Q-ball of frequency $\omega_Q = 0.75$ in the two channels $\sigma(\phi\to\phi)= \sigma_{-+}$, $\sigma(\phi\to\phi^*)= \sigma_{--}$.
Given the corresponding effective size of this Q-ball $R_Q \approx 3.3$, the cross section reaches the saturation for the angular momentum values  $\ell_{*} = R_Q \abs{k_-} \sim \{ 3, 11 \}$ for $\omega = \{2, 4 \}$ respectively.
This  cross section  in dimensionless units due to the variable redefinition above Eq.~\eqref{eq:1-field-action}. The correct scaling can be recovered by multiplying it by $\mu^{-2}$.
\label{fig:xsec-vs-l}
}
\end{figure}

\begin{figure}[H]
	\centering
	\includegraphics[scale=0.6]{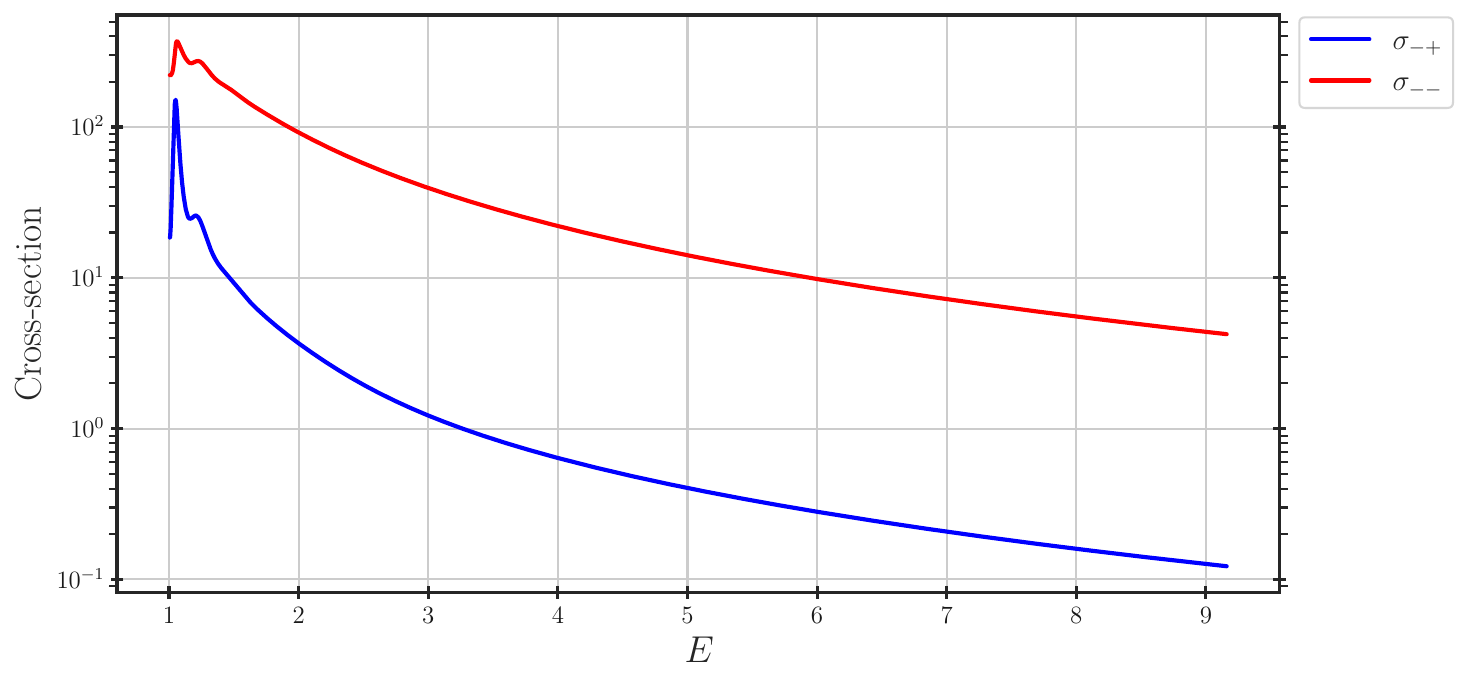}
  \caption{
	The total cross sections of incoming $-$ mode with energy $E$ scattering off a Q-ball of frequency $\omega_Q= 0.75$. 
  The red line represents the result for elastic channel (with outgoing mode $-$), while the blue line represents the inelastic one.
\label{fig:xsec-vsE1}
}
\end{figure}

\begin{figure}[H]
	\centering
	\includegraphics[scale=0.6]{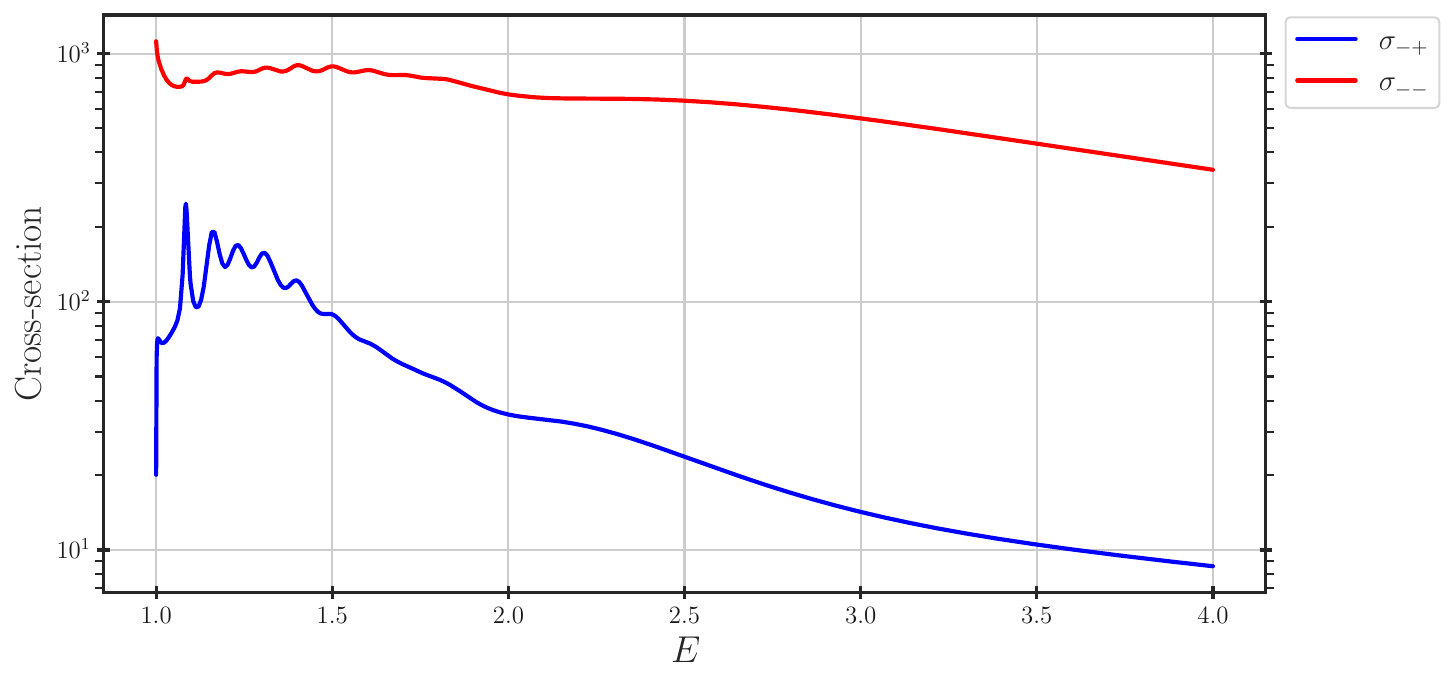}
  \caption{Same plot as of Fig.~\ref{fig:xsec-vsE1} for $\omega_Q=0.6$
  \label{fig:xsec-vsE2}
}
\end{figure}

From the symmetricity of the S-matrix: $S_\ell^{-+}(\omega) = S_\ell^{+-}(\omega)$, we can deduce the relation between the cross sections:
\bea
\sigma_{+ -}(E+\omega_Q)=\l[\frac{(E-\omega_Q)^2-1}{(E+\omega_Q)^2-1}\r]\sigma_{-+}(E-\omega_Q).
\eea
Note that similar equation for the $\sigma(\pm\pm)$ is not valid, as these expressions are sensitive to the phases of the $S_\ell$, which is not fixed using symmetricity and unitarity of the S-matrix.
We report our findings for the total cross section on the Fig.\ref{fig:xsec-vs-l},\ref{fig:xsec-vsE1},\ref{fig:xsec-vsE2}. We can see (Fig.\ref{fig:xsec-vs-l}) that as expected, the cross section saturates at   $\ell_{ *} \sim R_Q |k_\pm|$, where $R_Q$ is a typical Q-ball radius.
On the Fig.\ref{fig:xsec-vsE1}-\ref{fig:xsec-vsE2} we show the  cross section dependence on the energy of incoming particles for the Q balls with $\omega_Q=0.75$ and $\omega_Q=0.6$. We can see that there are some local minima in the total cross section dependence on energy. 
This effect is very similar to the appearance of zeros in the reflection coefficient  in 1-dimensional quantum mechanics, 
when we consider reflection/transmission of the plane wave over the potential well. Indeed it is well known that for certain values of 
energies the reflection coefficient is vanishing. In the case of the scattering over spherically symmetric 
well this leads the appearance of the zeros in the phase shifts for various values of orbital momentum, which 
leads to the local miniums in the total cross section once the sum over $l$ is performed. For the Q-balls the 
functions $U(r),W(r)$ controlling the dynamics of the perturbations indeed have a local minimum, acting as a 
potential well and depending on its size and location the appearance of 
the local minima becomes more  or less prominent.

In addition, we can look at the total cross-sections' dependence on the energy of the incoming particle or antiparticle

\section{Extra numerical results}
\label{appendix: Extra numerical results}

\subsection{Linear perturbation for two-field model}
\begin{figure}[H]
	\centering
\includegraphics[width={\textwidth}]{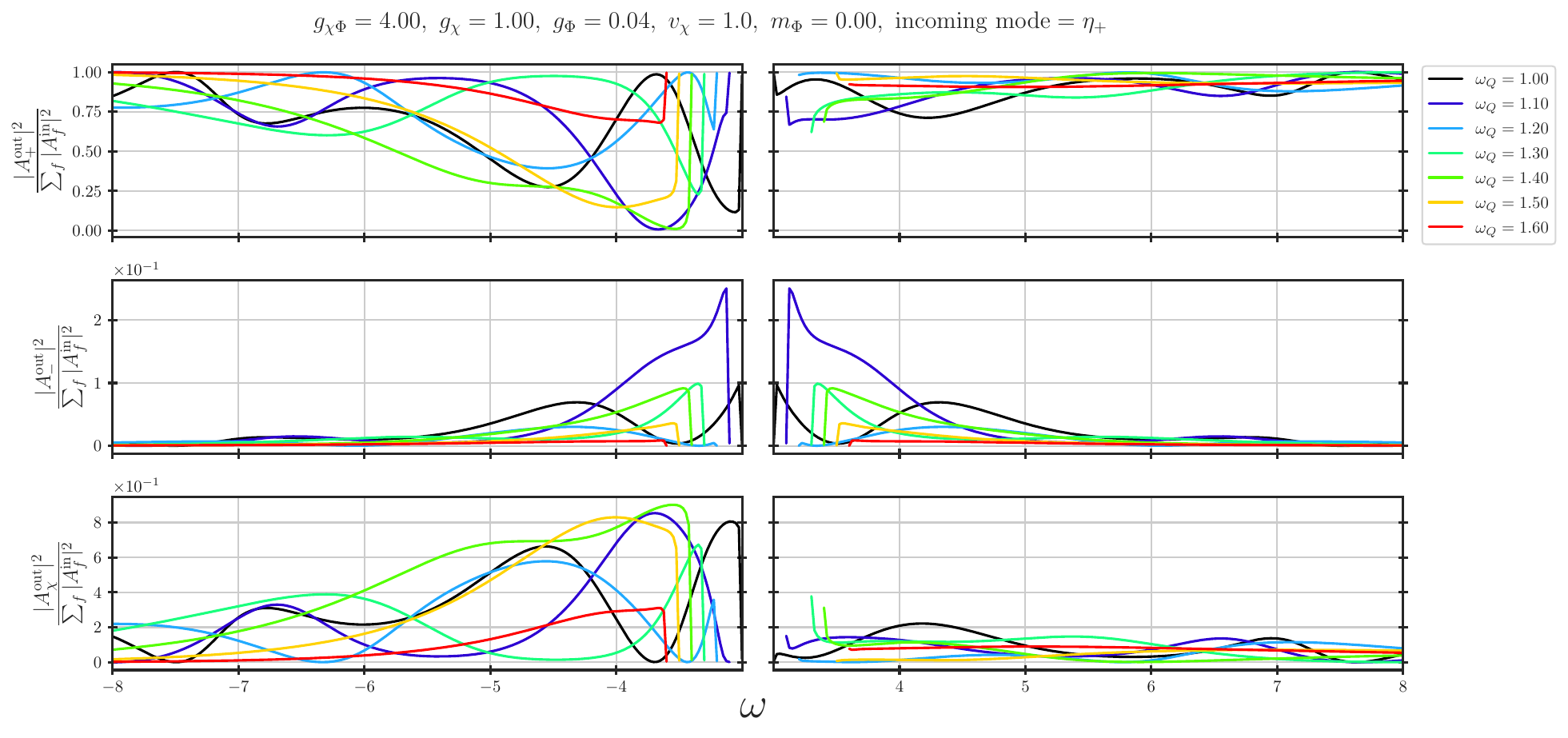}
	\caption{$\abs{A_f^{\rm out}}^2 / \abs{A_+^{\rm in}}^2$ for various $\omega_Q$. 
The enhancement/attenuation of squared amplitude for $+$ and $\chi$ modes exhibited greater fluctuations in the negative $\omega$ domain compared to the positive one.
Meanwhile, the amplification effect on the $-$ mode is even with respect to $\omega$.
}
	\label{fig: Aratio FLS model plus}
\end{figure}

\begin{figure}[H]
	\centering
	\includegraphics[width={\textwidth}]{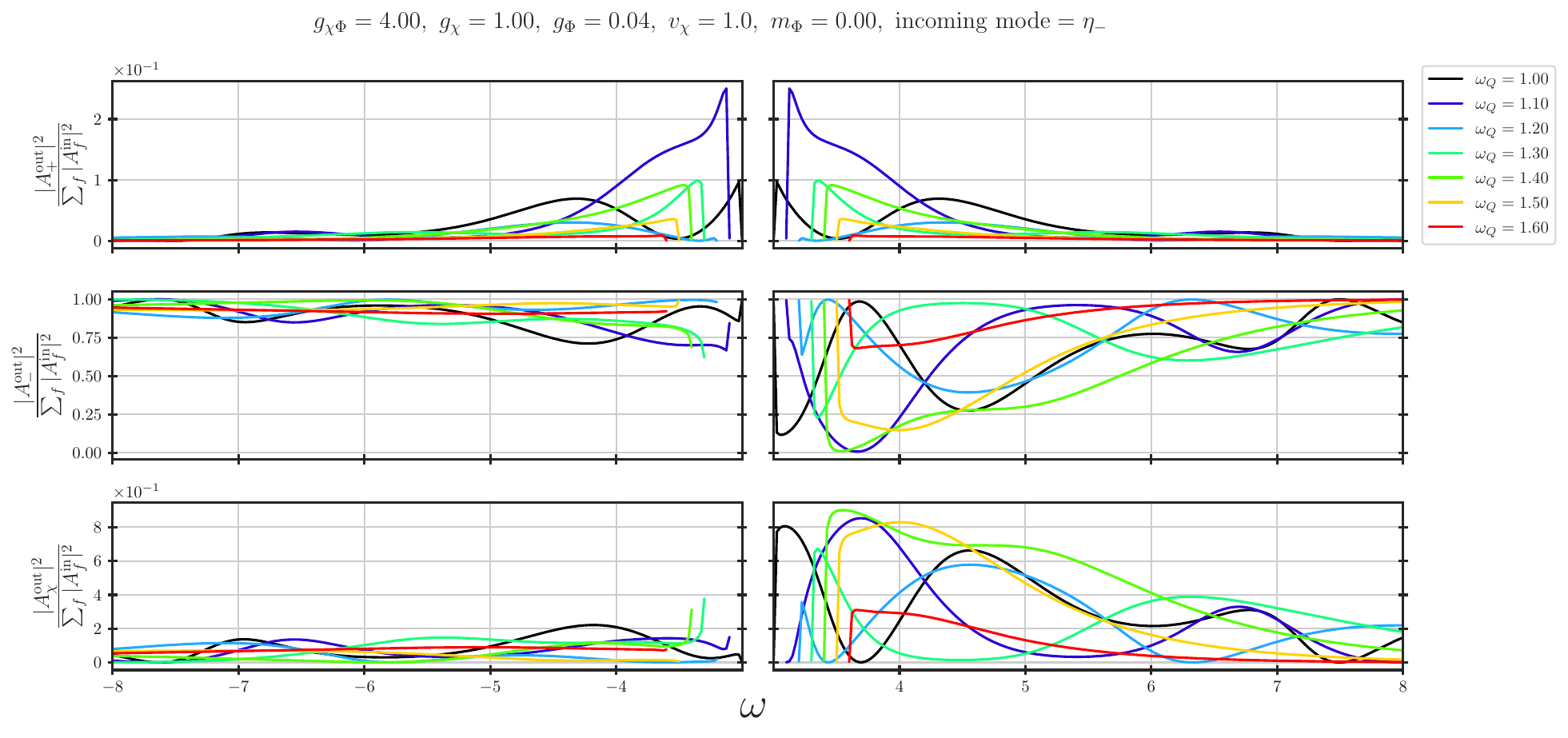}
	\caption{$\abs{A_f^{\rm out}}^2 / \abs{A_-^{\rm in}}^2$ for various $\omega_Q$. 
In comparison with Fig.~\ref{fig: Aratio FLS model plus}, one realizes the symmetries under $(\eta_\pm, \eta_\chi) \leftrightarrow (\eta_\mp^\ast, \eta_\chi^\ast)$.
\label{fig:fig17}
}
\end{figure}

\begin{figure}[H]
	\centering
	\includegraphics[width={\textwidth}]{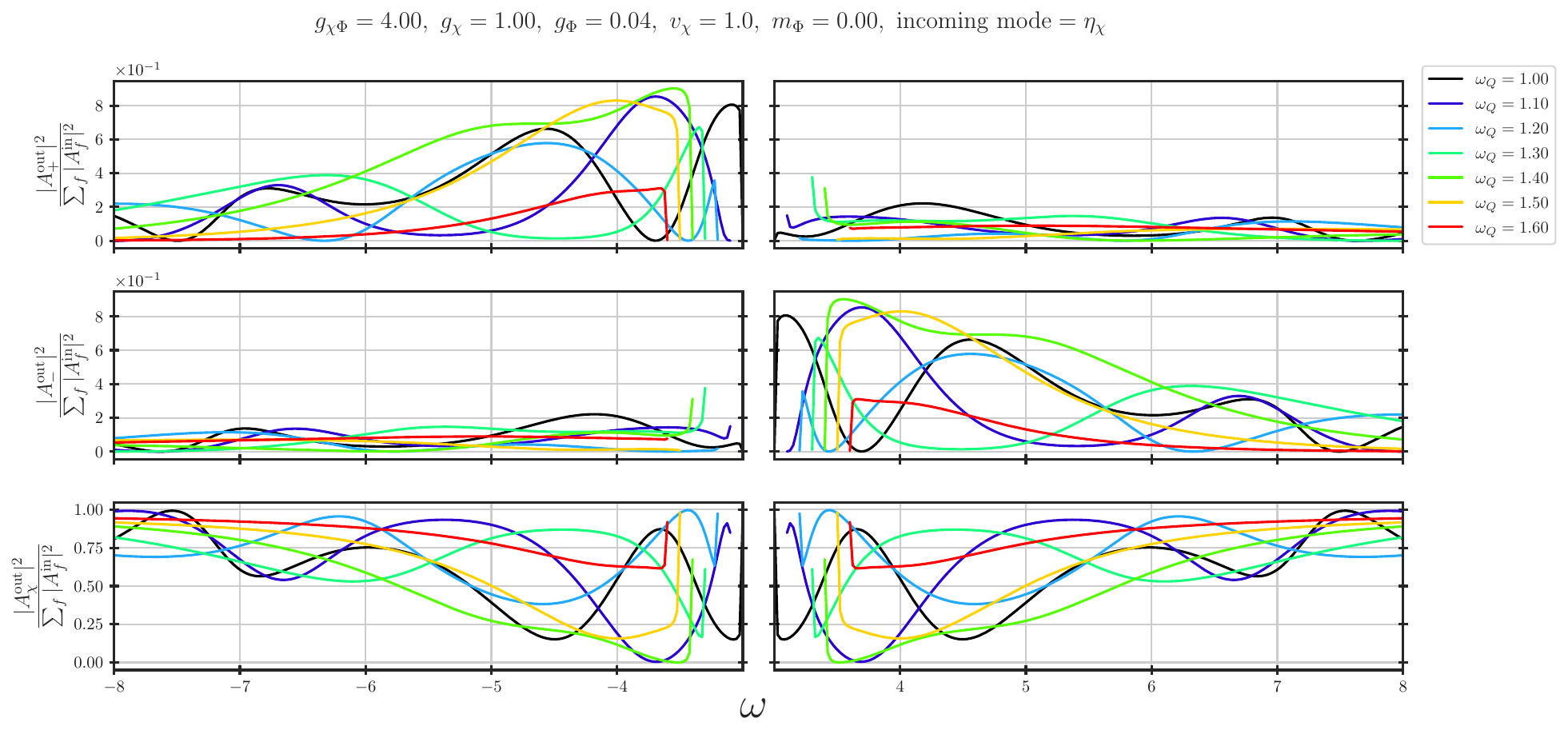}
	\caption{$\abs{A_f^{\rm out}}^2 / \abs{A_-^{\rm in}}^2$ for various $\omega_Q$. 
While the result for $f=\chi$ is an even function of $\omega$, we also have the symmetries between $f=+$ and $f=-$ under $\omega \rightarrow - \omega$.
\label{fig:fig18}
}
\end{figure}

\begin{figure}[H]
	\centering
	\includegraphics[width={\textwidth}]{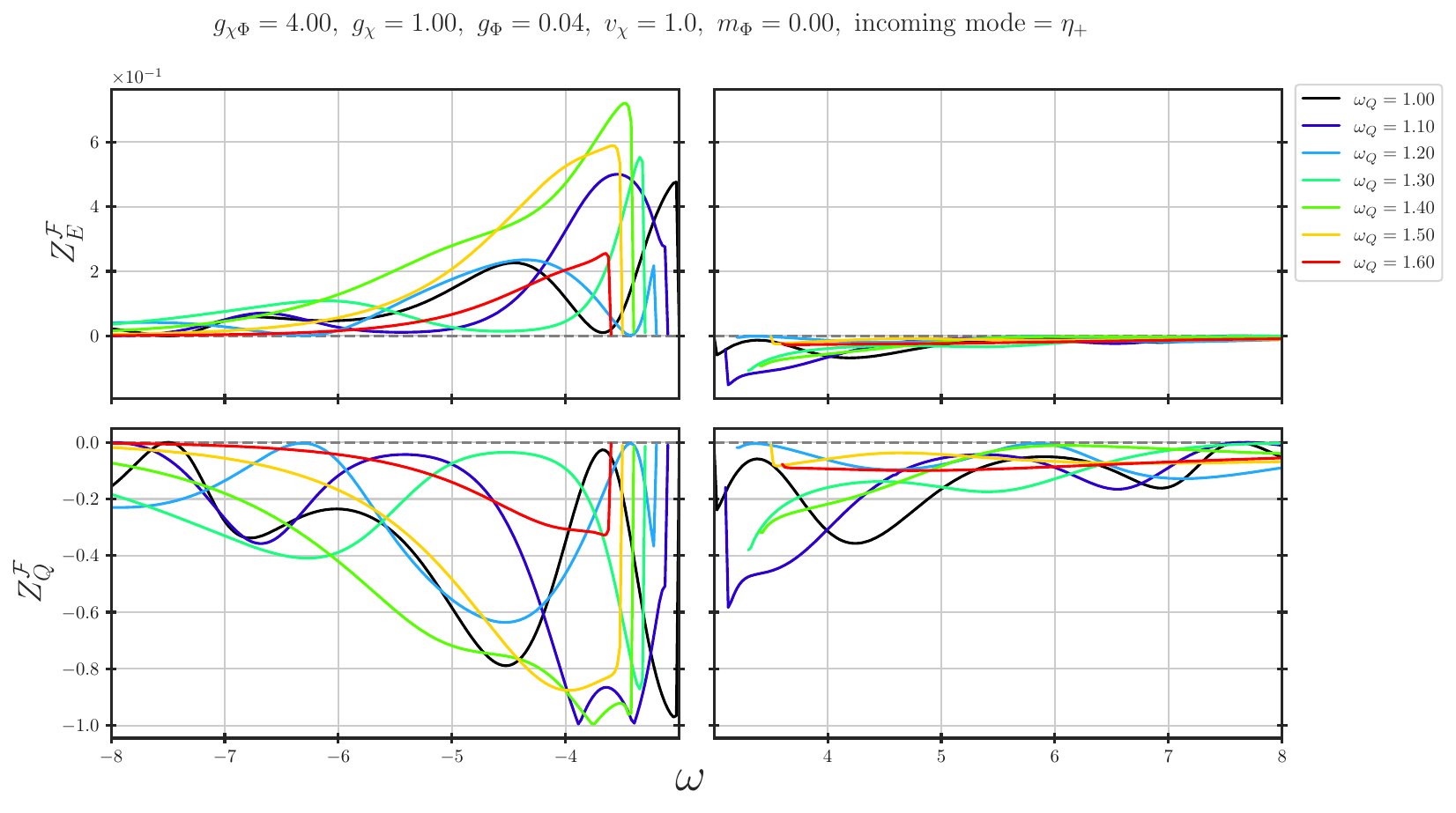}
	\caption{Energy and charge amplification factors as a function of incoming mode frequency with positively charged incoming mode.
The amplification/attenuation effects tends to be stronger with negative $\omega$ compared to the positive one.
The $Z_Q^{\mathcal{F}}$ is no longer symmetric under $\omega \rightarrow - \omega$ like the one field case due to the opening of a new scattering channel $+ \rightarrow \chi$.
\label{fig:fig19}
}
\end{figure}

\begin{figure}[H]
	\centering
	\includegraphics[width={\textwidth}]{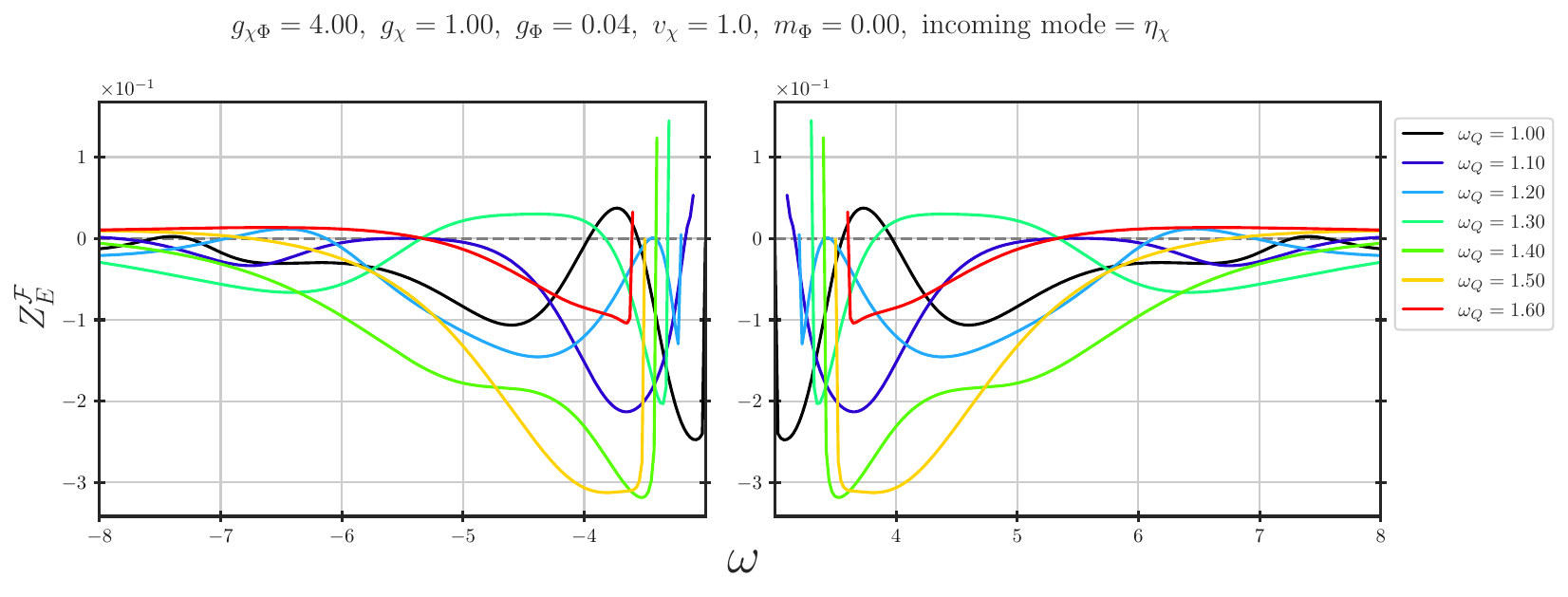}
	\caption{Energy amplification factors as a function of incoming mode frequency with neutral incoming mode.
We observe the symmetry pattern under $\omega \rightarrow - \omega$ in this case, which is a consequence of a property of the S-matrix discussed in previous appendix.
Since there is no net charge for the initial states, it is not well-defined to talk about the charge amplification.
\label{fig:fig20}
}
\end{figure}

In this section we report the 
ratios $\dfrac{\abs{A_f^{\rm out}}^2}{\abs{A_i^{\rm in}}^2} = \abs{S_{fi}}^2$, as well as the amplification factors $Z_{E, Q}$ for the Q-balls with various charges. The results are presented on the Fig.~\ref{fig: Aratio FLS model plus},\ref{fig:fig17},\ref{fig:fig18},\ref{fig:fig19},\ref{fig:fig20}. For all of these plots we have fixed the FLS model couplings to be equal to
$g_{\chi\Phi}=4, g_\chi=1, g_\Phi=0.04, v_\chi=1,m_\Phi=0$ and have varied the Q-ball charge ($\omega_Q$) and the incoming mode. All of the plots are done assuming the S-wave scattering. Symmetries proven in the Appendix.~\ref{app:symmetry} are evident.


One common pattern observed from these plots is that the quantities under considerations tends to be larger with small $\abs{\omega}$ and decay to their asymptotic values ($0$ or $1$ depending on the specific case) as $\abs{\omega}$ increases.
This implies the total elastic scattering of the incoming mode, or equivalently in the language of the S-matrix: $\lim_{\omega \rightarrow \pm \infty} S(\omega) = \mathds{1}$.

\subsection{Lattice simulation for two-field model}
\begin{figure}[H]
	\centering
	\begin{subfigure}{0.8\textwidth}
		\centering
		\includegraphics[width=\linewidth]{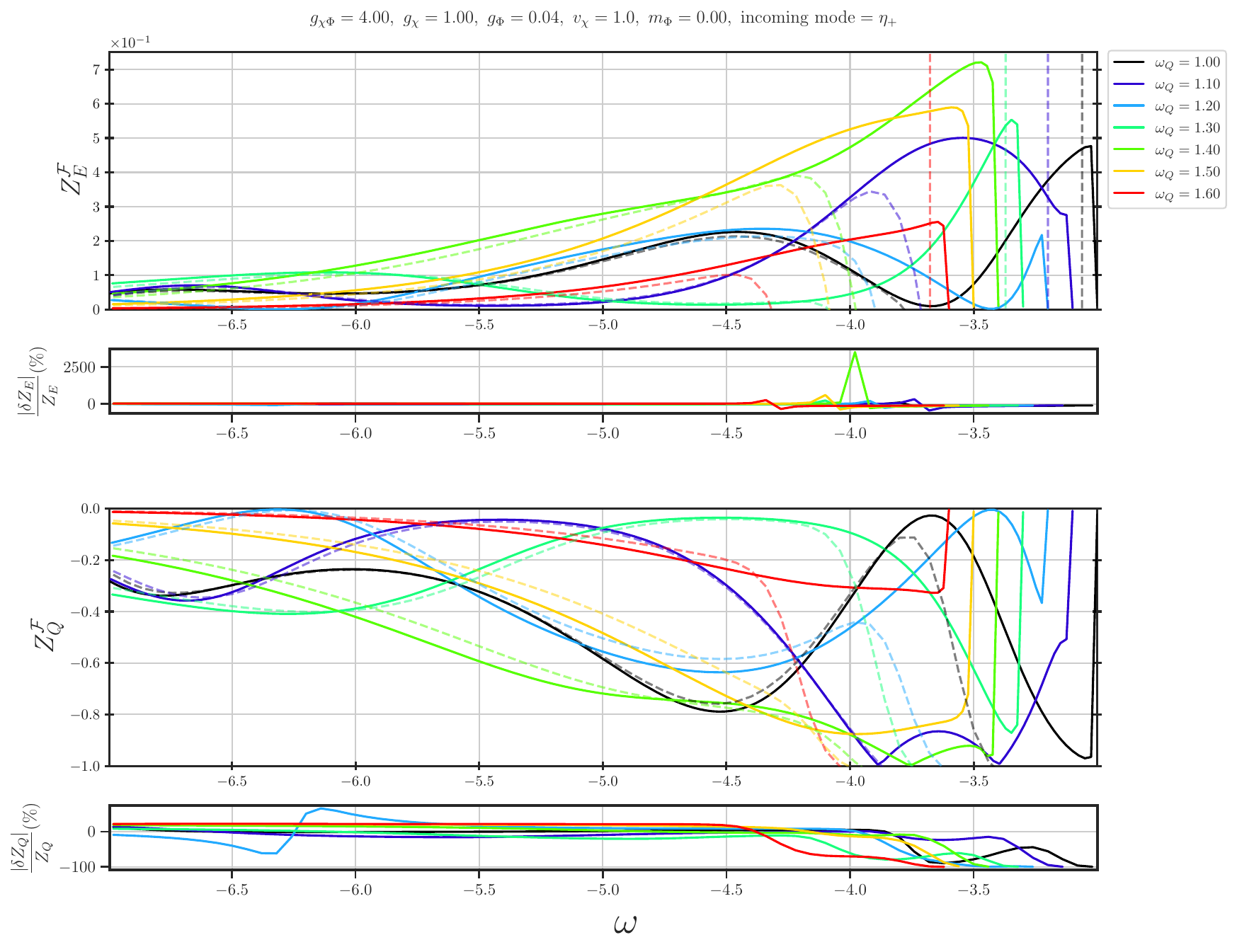}
	\end{subfigure}
	\begin{subfigure}{0.8\textwidth}
		\centering
		\includegraphics[width=\linewidth]{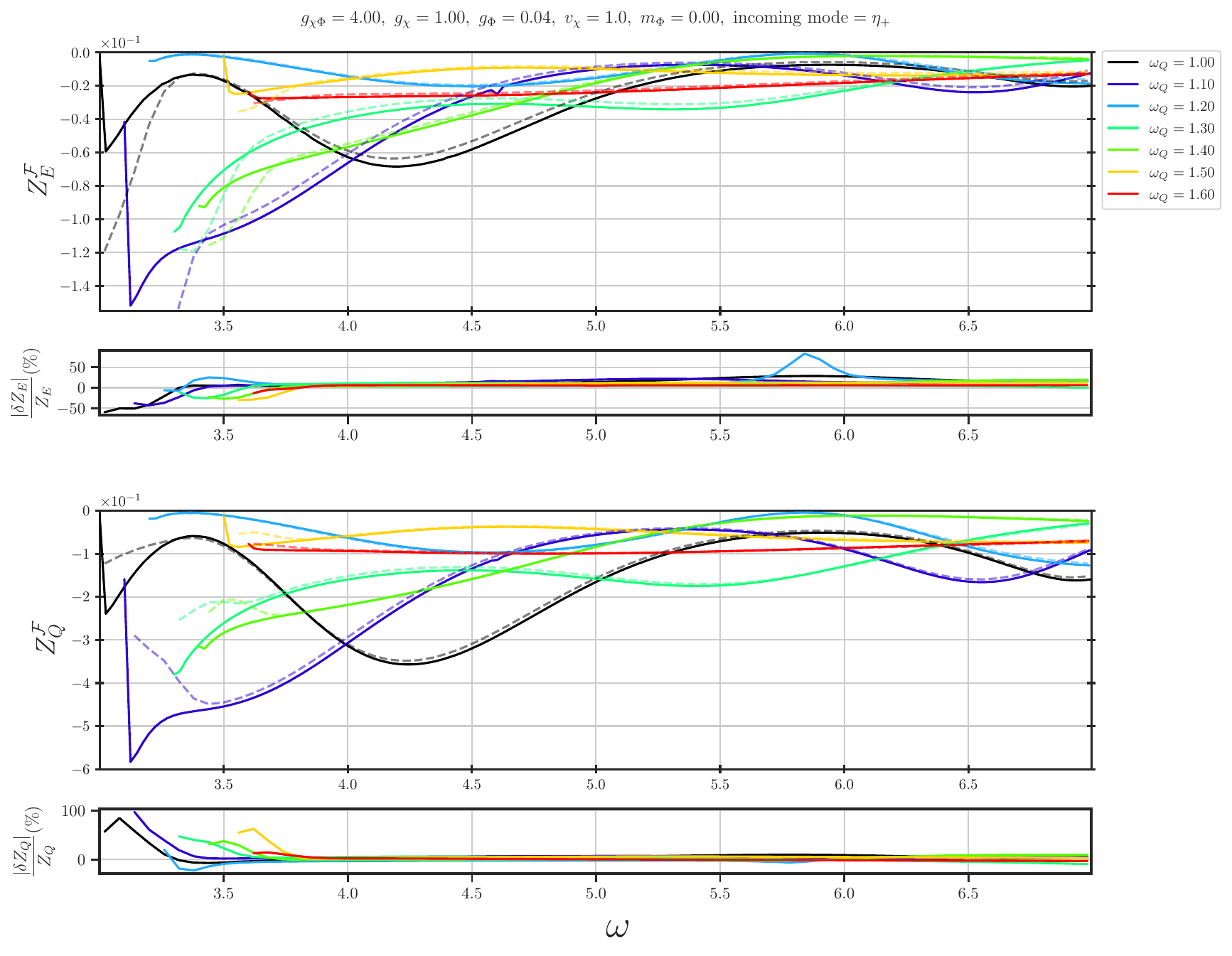}
	\end{subfigure}
	\caption{Comparison between amplification factors of linear regime (solid lines) with lattice results (dashed lines) on various Q-balls, with different colors corresponding to different $\omega_Q$.
Here we have kept the perturbation size fixed with $\sigma_r^\Phi = 10$ and $\delta_\Phi = 5 \times 10^{-4}$.
}
	\label{fig: FLS linear vs lattice}
\end{figure}

In this appendix we briefly discuss the numerical setup for the lattice 
calculations in the case of the two-field Q-ball. The numerical framework is the same as in the one field simulation with  4th-order finite difference discretization over space and Runge-Kutta 4th order integrator to evolve numerically over time.
The perturbed Q-balls initial conditions in $D=3$ with internal frequency $\omega_Q$ in FLS potential are
\begin{subequations}
	\begin{empheq}{align}
	&\Phi(0, r) = \Phi_Q(0, r) + \delta_\Phi \left( \dfrac{r_0}{r} \right) e^{- \frac{(r-r_0)^2}{2 (\sigma_r^\Phi)^2}} e^{-i k_\Phi r}, \\
	&\partial_t \Phi(0, r) = -i \omega_Q \Phi_Q(0, r) -i \omega_\Phi \delta_\Phi \left( \dfrac{r_0}{r} \right) e^{- \frac{(r-r_0)^2}{2 (\sigma_r^\Phi)^2}} e^{-i k_\Phi r}, \\
	&\chi(0, r) = \chi_Q(0, r) + \delta_\chi \left( \dfrac{r_0}{r} \right) e^{- \frac{(r-r_0)^2}{2 (\sigma_r^\chi)^2}} \cos(k_\chi r), \\
	&\partial_t \chi(0, r) = \delta_\chi \left( \dfrac{r_0}{r} \right)\omega e^{- \frac{(r-r_0)^2}{2 (\sigma_r^\chi)^2}} \sin(k_\chi r),
	\end{empheq}
\end{subequations}
where $k_\chi = \sqrt{\omega^2 - 8 g_\chi v_\chi^2}$ and $k_\Phi$ is either $k_+$ or $k_-$ depending on what incoming mode one wants to study, $k_\pm = \sqrt{\omega_\pm^2 - g_{\chi \Phi}^2 v_\chi^2 - m_\Phi^2}$.
For simplicity, we only consider one type of incoming mode for each scattering, hence turns on either $\delta_\Phi$ or $\delta_\chi$, but not both in the same simulation.
We perform a scan of amplification factors over various Q-balls and different values of incoming mode frequency $\omega$. 
We use the same set of $\omega_Q$ as previous results (check e.g the profiles for these Q-balls at Fig.~\ref{fig:qball-FLS}).
The final results are visualized on Fig.~\ref{fig: FLS linear vs lattice}.
\color{black}
For  these calculations we have kept the size of the perturbation fixed with the parameters $\sigma_r^\Phi = 10$ and $\delta_\Phi = 5 \times 10^{-4}$. We can see that for small values of $\omega$ the agreement between the lattice and linear description becomes worse. This is related to the larger wavelength of the perturbations, which at some point exceeds the width of the wave packet. Obviously the linear description cannot provide reliable results in this limit.

\bibliographystyle{JHEP}
{\footnotesize
\bibliography{biblio}}

\end{document}